\documentclass[11pt]{article}
\usepackage{amsfonts,amsmath,amssymb}
\usepackage{enumerate}
\usepackage{hyperref}
\usepackage{bbm}
\usepackage{nicefrac}
\usepackage[all]{xy}
\usepackage{graphicx}
\usepackage{bm}
\usepackage{makecell}
\usepackage[table]{xcolor}

\usepackage{upgreek}

\usepackage{booktabs}

\newcommand{\be}{\begin{equation}}
\newcommand{\ee}{\end{equation}}

\newcommand{\bea}{\begin{eqnarray}}
\newcommand{\eea}{\end{eqnarray}}

\newcommand{\bes}{\begin{subequations}}
\newcommand{\ees}{\end{subequations}}

\newcommand{\cN}{{\cal N}}

\newcommand{\tr}{\mbox{tr}}
\newcommand{\cZ}{{\cal Z}}

\newcommand{\et}{\frac{1}{3}}
\newcommand{\eft}{\frac{4}{3}}
\newcommand{\ett}{\frac{2}{3}}
\newcommand{\evt}{\frac{5}{3}}

\newcommand{\grp}[1]{\mathrm{#1}}
\newcommand{\rep}[1]{{\mathbf{#1}}}
\newcommand{\grU}{\grp{U}}
\newcommand{\grSU}{\grp{SU}}
\newcommand{\grSO}{\grp{SO}}

\def\sst#1{{\scriptscriptstyle #1}}

\def\0{{\sst{(0)}}}
\def\1{{\sst{(1)}}}
\def\2{{\sst{(2)}}}
\def\3{{\sst{(3)}}}
\def\4{{\sst{(4)}}}
\def\5{{\sst{(5)}}}
\def\6{{\sst{(6)}}}
\def\7{{\sst{(7)}}}
\def\8{{\sst{(8)}}}
\def\n{{\sst{(n)}}}

\def\cM{{{\cal M}}}

\newcommand{\cW}{{\cal W}}

\allowdisplaybreaks  

\usepackage{multirow}
\usepackage{rotating}

\newcommand{\ba}{\begin{align}}
\newcommand{\ea}{\end{align}}

\newcommand{\bse}{\begin{subequations}}
\newcommand{\ese}{\end{subequations}}

\allowdisplaybreaks

\let\OLDthebibliography\thebibliography
\renewcommand\thebibliography[1]{
  \OLDthebibliography{#1}
  \setlength{\parskip}{3.0pt}
}

\begin{document}

\makeatletter
\renewcommand{\theequation}{\thesection.\arabic{equation}}
\@addtoreset{equation}{section}
\makeatother

\begin{titlepage}

\begin{flushright}
IFT-UAM/CSIC-17-114
\end{flushright}

\vspace{25pt}

   \begin{center}
   \baselineskip=16pt

   \begin{Large}\textbf{
Spectrum universality properties \\[6pt]
of holographic Chern-Simons theories}
   \end{Large}

\vspace{25pt}
		
{\large  Yi Pang$^{1}$ ,\, Junchen Rong$^{2}$ \,and\,  Oscar Varela$^{1,3,4}$}
		
\vspace{30pt}

	\begin{small}

	  {\it $^{1}$ Max-Planck-Institut f\"ur Gravitationsphysik (Albert-Einstein-Institut), \\Am M\"uhlenberg 1, D-14476 Potsdam, Germany.  }

	\vspace{15pt}
	
	{\it $^{2}$  Fields, Gravity \& Strings, Center for Theoretical Physics of the Universe \\
	Institute for Basic Sciences, Daejeon 305-811, Korea}

	\vspace{15pt}
          
   {\it $^{3}$ Department of Physics, Utah State University, Logan, UT 84322, USA.}

	\vspace{15pt}
          
   {\it $^{4}$ Departmento de F\'\i sica Te\'orica and Instituto de F\'\i sica Te\'orica UAM/CSIC , \\
   Universidad Aut\'onoma de Madrid, Cantoblanco, 28049 Madrid, Spain}



	\end{small}

\vskip 50pt

\end{center}

\begin{center}
\textbf{Abstract}
\end{center}

\begin{quote}

We give a master formula for the spin-2 spectrum of a class of three-dimensional Chern-Simons theories at large $N$, with flavour group containing SU(3), that arise as infrared fixed points of the D2-brane worldvolume field theory and have AdS$_4$ duals in massive type IIA supergravity. We use this formula to compute the spin-2 spectrum of the individual theories, discuss its supermultiplet structure and, for an ${\cal N}=2$ theory in this class, the spectrum of protected operators with spin 2. We also show that the trace of the Kaluza-Klein graviton mass matrix on the dual AdS$_4$ solutions enjoys certain universality properties. These are shown to relate the class of AdS$_4$ massive IIA solutions under consideration to a similar class of AdS$_4$ solutions of $D=11$ supergravity with the same symmetries. Finally, for the ${\cal N}=2$ AdS$_4$ solution in this class, we study the entire spectrum at lowest Kaluza-Klein level and relate it to an analogue solution in $D=11$ supergravity.

\end{quote}

\vfill

\end{titlepage}

\tableofcontents



\section{Introduction}


The existence of Chern-Simons terms in addition to the usual Yang-Mills action in three dimensions renders the possible dynamics of gauge fields interacting with matter particularly rich in that number of dimensions. In this case, the Yang-Mills gauge coupling is irrelevant in the renormalisation group sense, and the Chern-Simons coupling is required by gauge invariance to be a quantised integer. For these reasons, the low energy dynamics of the Chern-Simons-Yang-Mills-matter system is dominated, precisely, by the Chern-Simons contributions. The fact that the Chern-Simons coupling is quantised and, therefore, unable to run with the energy scale, implies that Chern-Simons-matter theories may also enjoy conformal symmetry. Supersymmetry, and the tighter control on the dynamics that it encompasses, can be further added to the picture. Indeed, explicit Lagrangians involving SU$(N)$ gauge fields with Chern-Simons terms at level (that is, inverse coupling) $k$ and interacting matter can be constructed that are manifestly superconformal \cite{Schwarz:2004yj}. It is thus natural to enquire whether these field theories at strong 't Hooft coupling $\lambda  = N/k$ and large $N$ enjoy supergravity descriptions.

Some field theories of this type, like the Aharony-Bergman-Jafferis-Maldacena (ABJM) model \cite{Aharony:2008ug} on the M2-brane or the $\cN=2$ infrared fixed point \cite{Benna:2008zy} of a certain mass deformation of the former, are indeed known to  have supergravity duals. In those two cases, these correspond to the AdS$_4$ solutions of $D=11$ supergravity constructed by Freund-Rubin \cite{Freund:1980xh} and by Corrado, Pilch and Warner (CPW) \cite{Corrado:2001nv}, respectively. These and all other dual pairs of this type known until recently involved quiver field theories, rather than the simpler type of theories of \cite{Schwarz:2004yj} with a single gauge group. In fact, most of these simple super-Chern-Simons-matter theories \cite{Schwarz:2004yj} {\it do not} have weakly-coupled AdS$_4$ supergravity duals. The reason is that their spectrum typically contains light operators with unbounded spin \cite{Minwalla:2011ma}, rather than only operators of at most spin 2. Furthermore, the spectrum of these theories tends to exhibit exponential growth at large $\lambda$ \cite{Minwalla:2011ma}, rather than the polynomial-type behaviour that a dual Kaluza-Klein (KK) description would predict. 

While the results of \cite{Minwalla:2011ma} rule out the existence of large-radius supergravity duals for most of the simple superconformal field theories in the class of \cite{Schwarz:2004yj}, they still leave a handful of cases with very specific matter content and interactions open to a holographic supergravity interpretation. Recently, this supergravity description has indeed been found \cite{Guarino:2015jca}. These particular superconformal Chern-Simons theories can be engineered as infrared phases \cite{Guarino:2015jca,Guarino:2016ynd} of the field theory defined on a stack of $N$ D2-branes in flat space, three-dimensional $\cN=8$ SU$(N)$ super-Yang-Mills, in the presence of a non-vanishing Romans mass, $\hat F_\0$. The latter is holographically identified with the Chern-Simons level, $\hat F_\0 = k$, similarly to \cite{Gaiotto:2009mv}. Accordingly, infrared field theories of this type with $\cN$ supersymmetries and flavour group $G$ contained in the SO(7) R-symmetry of the ultraviolet $\cN=8$ super-Yang-Mills, are dual  to $\cN$--supersymmetric AdS$_4 \times S^6$ solutions of massive type IIA supergravity \cite{Romans:1985tz}, equipped with a metric and background fluxes on the internal six-sphere $S^6$ that are locally invariant under $G \subset \textrm{SO}(7)$. Precise dual pairs with $\cN =2 , \textrm{SU}(3) \times \textrm{U}(1)$ \cite{Guarino:2015jca} and $\cN =3 , \textrm{SO}(4)$ \cite{Pang:2015vna} symmetry have been identified. Massive IIA $\cN=1$ solutions in the same class, which still await a precise field theory interpretation, have also been constructed with $\textrm{G}_2$ \cite{Behrndt:2004km, Varela:2015uca} and $\textrm{SU}(3)$ \cite{Varela:2015uca} symmetry\footnote{Previously known numerical solutions in the same or a similar class include \cite{Petrini:2009ur,Lust:2009mb}. A different class of $\cN=1$ AdS$_4$ solutions of massive type IIA has been found recently in \cite{Apruzzi:2015wna,Rota:2015aoa}.}. Further aspects of these new AdS$_4$/CFT$_3$ dualities have now been developed in \cite{Fluder:2015eoa,Hosseini:2016tor,Araujo:2016jlx,Guarino:2017eag,Guarino:2017pkw,Araujo:2017hvi,Azzurli:2017kxo,Hosseini:2017fjo,Benini:2017oxt}. 

In this paper, we set out to study holographically an important aspect of these superconformal Chern-Simons-matter theories at large $N$: their spectrum of single-trace operators. For any of these theories with supersymmetry $\cN$ and flavour group $G \subset \textrm{SO}(7)$, the spectrum is organised in representations of $G$ and supermultiplets of the $\cN$--extended three-dimensional superconformal group, OSp$(4|\cN)$, with states of up to spin 2. Partial results are already available. The protected spectrum of the $\cN =3 , \textrm{SO}(4)$ field theory has been investigated directly from the field theory \cite{Minwalla:2011ma}. Also for this $\cN=3$ theory, the large-$N$ spectrum of operators, not necessarily protected, with spin 2 has been determined from the supergravity \cite{Pang:2015rwd}. The holographic determination of the single-trace operators of these field theories involves the calculation of the entire KK spectrum of massive type IIA supergravity on the corresponding AdS$_4 \times S^6$ solutions. This appears to be prohibitively difficult, given the relatively small isometry groups $G$ and the presence of background supergravity fluxes. For simplicity, in this paper we will rather focus in the particular subsector of the spectra containing operators of spin 2, for the $\cN=2$ field theory of \cite{Guarino:2015jca} and the (still unknown) $\cN=1$ field theories of \cite{Behrndt:2004km, Varela:2015uca}. This entails the calculation of the spectrum of KK gravitons about the massive type IIA AdS$_4 \times S^6$ solutions of \cite{Guarino:2015jca,Behrndt:2004km, Varela:2015uca}. Previous calculations of massive KK graviton spectra in related contexts include \cite{Klebanov:2008vq,Klebanov:2009kp,Bachas:2011xa,Richard:2014qsa,Passias:2016fkm}.

While the determination of the entire KK towers of fields of all spin $s < 2$ on AdS$_4$ is a very difficult problem, we nevertheless do have access to a full sector of the KK spectrum: the slice containing all $s \leq 2$ modes that lie at the bottom, $n=0$, of the KK towers of massive IIA supergravity on any of the relevant AdS$_4 \times S^6$ solutions. The reason is that this class of D2-brane AdS$_4$/CFT$_3$ dualities \cite{Guarino:2015jca} is very special in that it belongs to the distinguished class of holographic dualities that enjoy a (partial) large--$N$ effective description in terms of maximally supersymmetric gauged supergravity. Well-known examples with this remarkable property include the M2-brane AdS$_4$/CFT$_3$ examples of \cite{Aharony:2008ug,Benna:2008zy} and the D3-brane AdS$_5$/CFT$_4$ cases of \cite{Maldacena:1997re,Freedman:1999gk}, whose partial effective description is respectively provided by $D=4$ $\cN=8$ (electrically) gauged SO(8) supergravity \cite{deWit:1982ig} and $D=5$ $\cN=8$ SO(6)--gauged supergravity \cite{Gunaydin:1984qu}. In the case at hand, the relevant $D=4$ $\cN=8$ supergravity has a dyonically gauged (in the sense of \cite{Dall'Agata:2012bb,Dall'Agata:2014ita,Inverso:2015viq}) $\textrm{ISO}(7) \equiv \textrm{CSO}(7,0,1) \equiv \textrm{SO}(7) \ltimes \mathbb{R}^7$ gauge group. 

This gauged supergravity has been explicitly constructed \cite{Guarino:2015qaa} and shown to arise upon consistent KK truncation of massive type IIA supergravity on $S^6$ \cite{Guarino:2015jca,Guarino:2015vca}. Its (AdS) vacua \cite{DallAgata:2011aa,Gallerati:2014xra,Guarino:2015qaa} (see table 1 of the latter reference for a summary of the known ones) uplift on $S^6$ to the AdS$_4 \times S^6$ solutions  \cite{Guarino:2015jca,Pang:2015vna,Behrndt:2004km,Varela:2015uca}  of massive type IIA supergravity discussed above. As in the M2 \cite{Aharony:2008ug,Benna:2008zy} and D3-brane cases \cite{Maldacena:1997re,Freedman:1999gk}, the $\cN=8$ supergravity captures and reconstructs the full, non-linear dynamics of the modes that arise upon linearisation of massive IIA supergravity around those AdS$_4 \times S^6$ solutions at the $n=0$ bottom of the KK towers. For the maximally supersymmetric cases on the M2 \cite{Aharony:2008ug} and D3 \cite{Maldacena:1997re} branes, the $\cN=8$ supergravity modes are dual to the maximally supersymmetric stress-energy tensor supermultiplet. For M2 and D3 brane cases with less supersymmetry \cite{Benna:2008zy,Freedman:1999gk}, the $\cN=8$ modes split into the relevant stress-energy supermultiplet and other matter multiplets. The present D2-brane examples are like the latter cases, because all vacua of dyonic $\textrm{ISO}(7)$ supergravity spontaneously break $\cN=8$ supersymmetry (and SO(7) symmetry). 

The vacua of $D=4$ $\cN=8$ gauged supergravities enjoy a curious universality property: vacua of {\it different} gauged supergravities that preserve {\it the same} symmetry group $G$ (regarded as a subgroup of their respective gauge groups) tend to exhibit the same mass spectrum within their corresponding supergravities\footnote{This appears to be the case at least for gauged supergravities with a higher-dimensional origin. The three different SO(4)--invariant vacua \cite{Borghese:2013dja} of electric SO(8) supergravity \cite{deWit:1982ig} have different mass spectra, but their actual symmetry is also different: they preserve distinct discrete symmetries in addition to each SO(4). The non-supersymmetric SU(3)--invariant point \cite{Borghese:2012zs} of dyonic SO(8) supergravity \cite{Dall'Agata:2012bb} does evade this rule, but the latter theory does not admit a higher-dimensional origin \cite{deWit:2013ija,Lee:2015xga,Inverso:2017lrz}.}, see \cite{Dall'Agata:2012bb,DallAgata:2011aa,Borghese:2012zs}. This is the case even if the common residual symmetry $G$ is embedded differently in their respective gauge groups and ultimately in E$_{7(7)}$. For example, electrically-gauged SO(8) supergravity \cite{deWit:1982ig} and dyonic ISO(7) supergravity \cite{Guarino:2015qaa} both have an $\cN=2$ AdS critical point with $\textrm{SU}(3) \times \textrm{U}(1)$ residual symmetry \cite{Warner:1983vz,Guarino:2015jca,Guarino:2015qaa} with the same mass spectrum \cite{Nicolai:1985hs,Guarino:2015qaa} within their respective $\cN=8$ theories. For $\cN=8$ supergravities that descend from higher dimensions, this type of vacua thus display a universal mass spectrum at lowest KK level. In other words, the $n=0$ KK mass spectrum in these cases is insensitive to the precise higher-dimensional origin and compactification manifold. Two questions therefore arise.

The first question is whether the universality of the $n=0$ KK mass spectrum is lost at higher, $n>0$ KK levels. Intuition dictates that this should be the case --a full KK spectroscopy analysis should be able to tell these compactifications apart. We address this question for the CPW AdS$_4$ solution \cite{Corrado:2001nv} of $D=11$ supergravity and the massive type IIA AdS$_4$ solution of \cite{Guarino:2015jca}. These correspond respectively to the $S^7$ \cite{deWit:1986iy} and $S^6$ \cite{Guarino:2015jca,Guarino:2015vca} uplifts of the $\cN=2$ $\textrm{SU}(3) \times \textrm{U}(1)$--invariant critical points of electric SO(8) \cite{Warner:1983vz} and dyonic ISO(7) \cite{Guarino:2015jca,Guarino:2015qaa} $\cN=8$ supergravity. The dual SU(3)--flavoured $\cN=2$ field theories have been discussed in \cite{Benna:2008zy} and \cite{Guarino:2015jca}. Fortunately, we do not need to perform a full KK analysis for both solutions \cite{Corrado:2001nv,Guarino:2015jca} as the spin-2 subsector suffices to draw conclusions. The respective towers of massive KK gravitons, computed for the CPW solution in \cite{Klebanov:2009kp} and in this paper for the solution of \cite{Guarino:2015jca}, do differ already at first KK level. The spectrum of dual spin-2 operators thus differs too.

Somewhat surprisingly, however, universality still persists in a milder capacity. We introduce the infinite-dimensional graviton mass matrix for these solutions and diagonalise it at fixed KK level $n$, for all $n =1,2, \ldots$ While the individual eigenvalues are indeed different, leading to different graviton spectra for both solutions \cite{Corrado:2001nv} and \cite{Guarino:2015jca}, we find that the graviton mass matrix traces match KK level by KK level. See equation (\ref{tracerelation}) for a more precise statement. Thus, while the strong, eigenvalue by eigenvalue, universality of the $n=0$ KK mass spectra of these solutions is lost at higher KK levels $n>0$, a softer form of universality still persists at the level of the mass matrix traces, at least in the spin-2 sector. Our analysis also reveals a related universality property. For all AdS$_4 \times S^6$ solutions of massive IIA known to uplift from critical points of dyonic ISO(7) supergravity, the trace of the graviton mass matrix at fixed KK level $n$ is given by a universal polynomial in $n$, times an overall constant that depends on the individual solution.

The second question is whether the spectrum of dual operators differs already at lowest KK level, even if the supergravity masses are identical at that level. This is conceivable because different conformal dimensions may be related to the same supergravity mass. 
 We tackle this question by focusing on the complete $0 \leq s \leq 2$ spectrum of the $\cN=2$ vacua  \cite{Warner:1983vz,Guarino:2015jca} within the electric SO(8) and the dyonic ISO(7) supergravities. The OSp$(4|2) \times \textrm{SU}(3)$ supermultiplet structure of the spectrum in the former case was elucidated in \cite{Nicolai:1985hs} and revisited more recently in \cite{Klebanov:2008vq,Klebanov:2009kp}. Here, we perform the analogue analysis for the $\cN=2$ point of dyonic ISO(7) supergravity. We find an identical supermultiplet structure, with a subtle difference.  The R-charges and conformal dimensions are the same in both cases, except for two SU(3) sextet hypermultiplets which exhibit different assignments. This difference can be put down to different U(1) R-symmetries being preserved. In fact, two possible choices \cite{Nicolai:1985hs}, named scenarios I and II in  \cite{Klebanov:2008vq,Klebanov:2009kp}, were noted to be in principle possible for this U(1). The spectrum of the $\cN=2$ point \cite{Warner:1983vz} of SO(8) supergravity realises scenario I \cite{Nicolai:1985hs,Klebanov:2008vq,Klebanov:2009kp}. We find that the spectrum of the $\cN=2$ point of dyonic ISO(7) supergravity realises scenario II.

\section{Massive gravitons with at least SU(3) symmetry} \label{sec:KKGravitonsSU3}


\subsection{Background geometry} \label{subsec:Setup}

We are interested in (Einstein frame) type IIA geometries of the form
\begin{eqnarray} \label{10Dgeom}
d\hat{s}^2_{10} = e^{2A (y) } \Big[ \big(  \bar{g}_{\mu\nu} (x) + h_{\mu\nu} (x,y) \big) d x^\mu dx^\nu +d\bar{s}_6^2 (y) \, \Big] \; ,
\end{eqnarray}
where $(x,y)$ collectively denote the external and internal coordinates, respectively. The metrics $\bar{g}_{\mu\nu}$, $\mu , \nu = 0 , \ldots, 3$, and $d\bar{s}_6^2$ correspond to the background geometry and  $ h_{\mu\nu} $ to a perturbation over the background. For the external background geometry we take four-dimensional, unit radius anti-de Sitter space, $ \bar{g}_{\mu\nu} d x^\mu dx^\nu = ds^2 ( \textrm{AdS}_4)$. The warp factor $e^{2A}$ and internal  geometry $d\bar{s}_6^2$ will be specified shortly. We write the perturbation in the factorised form
\begin{eqnarray} \label{PerturbExpansion}
h_{\mu\nu} (x,y) = h_{\mu\nu}^{[\textrm{tt}]} (x) \, Y(y) \; , 
\end{eqnarray}
where $Y(y)$ is a function on the internal six-dimensional space and  $ h_{\mu\nu}^{[\textrm{tt}]} $ is transverse, $\bar{\nabla}^\mu  h_{\mu\nu}^{[\textrm{tt}]} = 0 $, with respect to the Levi-Civita connection corresponding to $\bar{g}_{\mu\nu}$, and traceless, $\bar{g}^{\mu\nu} h_{\mu\nu}^{[\textrm{tt}]} = 0$. The perturbation is taken to satisfy the Fierz-Pauli equation,
\begin{eqnarray} \label{FPeq0}
\overline{\Box} \, h_{\mu\nu}^{[\textrm{tt}]}  = \big( M^2 L^2 -2 \big) \,  h_{\mu\nu}^{[\textrm{tt}]} \; , 
\end{eqnarray}
for a graviton of mass $M^2$ propagating on the background AdS$_4$ space. Here, $L$ is the radius of AdS$_4$ introduced by the warping $e^{2A}$ in (\ref{10Dgeom}) (see (\ref{warping})), and will be defined in (\ref{VSU3}). Under these assumptions, the linearised ten-dimensional Einstein equations devolve into the following second-order differential equation for $Y(y)$ \cite{Bachas:2011xa}:
\begin{eqnarray} \label{FPeq}
- \frac{e^{-8A}}{\sqrt{\bar{g}}} \partial_m \Big( e^{8A} \sqrt{\bar{g}} \,  \bar{g}^{mn} \partial_n \Big) Y = M^2 L^2  \, Y  \; , 
\end{eqnarray}
where $\bar{g}^{mn}$, $m,n = 1 , \ldots , 6$, and $\bar{g}$ respectively are the inverse metric components and the determinant of the internal line element $d\bar{s}_6^2$ in (\ref{10Dgeom}). 

For the internal background geometry we take the following family of six-dimensional metrics \cite{Varela:2015uca}
{\setlength\arraycolsep{2pt}
\begin{eqnarray} \label{KKSU3sectorinIIA}
d\bar{s}_{6}^2 =  L^{-2} g^{-2} \,   \Big[ \;  e^{-2\phi+\varphi}   X^{-1}  d\alpha^2      +    \sin^2 \alpha  \Big( \Delta_1^{-1} ds^2 ( \mathbb{CP}^2 ) + X^{-1} \Delta_2^{-1} (d\psi + \sigma)^2  \Big) \Big] \; . 
\end{eqnarray}
}Here, $\alpha$ and $\psi$ are angles with ranges
\begin{eqnarray} \label{anglerange}
0 \leq \alpha \leq \pi \; ,   \qquad 
0 \leq \psi \leq 6 \pi \; ,
\end{eqnarray}
$ds^2 ( \mathbb{CP}^2 )$ is the Fubini-Study metric on the complex projective plane, normalised so that the Ricci tensor equals six times the metric, and $\sigma$ is a one-form potential for the K\"ahler form  $\bm{J} $ on $\mathbb{CP}^2$, normalised as $ d \sigma = 2  \bm{J} $. The metrics (\ref{KKSU3sectorinIIA}) depend on five parameters, $\varphi$, $\chi$, $\phi$, $\zeta$, $\tilde{\zeta}$ through the combinations\footnote{We define the combination of parameters $Y$ following \cite{Guarino:2015qaa,Varela:2015uca}. Here and in the formulae below, this should not be confused with the eigenfunction $Y$ defined in (\ref{PerturbExpansion}), (\ref{FPeq}).}
\begin{eqnarray} \label{eq:combis}
& X \equiv   1 + e^{2\varphi} \chi^2 \; , \qquad 
Y \equiv  1 + \tfrac14 e^{2\phi}  (\zeta^2 + \tilde \zeta^2) \; ,    \nonumber  \\[4pt]
& \Delta_1 \equiv   e^{\varphi} \, Y \sin^2 \alpha  + e^{2\phi-\varphi} X \cos^2 \alpha \; ,   \qquad
\Delta_2 \equiv   e^{\varphi}  \sin^2 \alpha  + e^{2\phi-\varphi}   \cos^2 \alpha \; . 
\end{eqnarray}
These parameters take values on the six-dimensional manifold\footnote{ The sixth parameter in (\ref{scalarmanifold}), called $a$ in \cite{Varela:2015uca}, does not enter the line element (\ref{KKSU3sectorinIIA}). }
\begin{eqnarray} \label{scalarmanifold}
\frac{\textrm{SU}(1,1)}{\textrm{U}(1)} \times \frac{\textrm{SU}(2,1)}{\textrm{SU} (2) \times \textrm{U}(1)} \; .
\end{eqnarray}
The constant $g$ in (\ref{KKSU3sectorinIIA}) is non-vanishing, and $L$ is defined as $L^2 = -6 \, V^{-1}$, where $V$ is the following function on (\ref{scalarmanifold}),
\begin{equation}
\label{VSU3}
\begin{array}{lll}
V &=& \frac{1}{2} \, g^{2} \left[ e^{4 \phi -3 \varphi } \big(1+e^{2\varphi} \chi^2 \big)^3 -12  \, e^{2 \phi -\varphi }  \big(1+e^{2\varphi} \chi^2 \big) -24 \, e^{\varphi} \right.
\\[6pt]
&& \left. \qquad +  \tfrac34 \, e^{4\phi +\varphi} \big( \zeta^2 + \tilde \zeta^2 \big)^2   \big(1+3 \, e^{2\varphi} \chi^2 \big)+ 3\, e^{4\phi +\varphi} \big( \zeta^2 + \tilde \zeta^2 \big) \chi^2   \big(1+e^{2\varphi} \chi^2 \big) \right.
\\[6pt]
&& \left. \qquad  -3 \, e^{2\phi +\varphi} \big( \zeta^2 + \tilde \zeta^2 \big)   \big(1-3 \, e^{2\varphi} \chi^2 \big)   \right] -    \tfrac12 \, g \, m \, \chi  \, e^{4 \phi+3 \varphi  } \left( 3\big( \zeta^2 + \tilde \zeta^2 \big) + 2\chi ^2 \right)  \\[6pt]
&& + \,  \frac{1}{2} \, m^2 \, e^{4 \phi+3 \varphi  }  \ .
\end{array}
\end{equation}
It depends on $g$ and on a further constant $m$, which is also non-vanishing. Finally, the warp factor in (\ref{10Dgeom}) is 
\begin{eqnarray} \label{warping}
e^{2A} = e^{\frac18 (2\phi-\varphi)} X^{1/4}  \Delta_1^{1/2}  \Delta_2^{1/8}  \, L^{2} \; . 
\end{eqnarray}
For all values of the parameters and with the periodicities (\ref{anglerange}), the local line element (\ref{KKSU3sectorinIIA}) extends globally into a smooth geometry on $S^6$ \cite{Varela:2015uca}.

The internal geometry (\ref{KKSU3sectorinIIA}) corresponds to the uplift \cite{Varela:2015uca} of the dynamical SU(3)--invariant sector of $D=4$ $\cN=8$ dyonically-gauged ISO$(7)$ supergravity \cite{Guarino:2015qaa}, as follows from the general consistent truncation of massive type IIA supergravity on the six-sphere \cite{Guarino:2015jca,Guarino:2015vca}. In general, the parameters valued in (\ref{scalarmanifold}) correspond to the four-dimensional scalar fields that preserve the SU(3) subgroup of the ISO(7) gauge group of the supergravity, $g$ and $m$ respectively are the electric and magnetic gauge couplings, and the function $V$ is the SU(3)--invariant scalar potential \cite{Guarino:2015qaa}. At the critical points of the scalar potential (\ref{VSU3}), recorded in table 3 of \cite{Guarino:2015qaa}, the $D=4$ scalars become fixed to the corresponding constant vacuum expectation values (vevs), and the geometry (\ref{10Dgeom}), (\ref{KKSU3sectorinIIA}) with $h_{\mu\nu} = 0$ becomes the warped product of AdS$_4$ and a topological $S^6$. These geometries are supported by IIA fluxes \cite{Varela:2015uca}, whose expressions will not be needed in what follows. These solutions of massive type IIA supergravity are dual to large-$N$ Chern-Simons field theories with a single gauge group and flavour group containing SU(3) \cite{Guarino:2015jca}. Note that the $L$ that we are using here is different than the $L$'s defined for each AdS$_4 \times S^6$ solution on a case-by-case basis in \cite{Guarino:2015jca,Varela:2015uca}. 

We want to compute the spectrum of KK gravitons $h_{\mu \nu}$ above these AdS$_4$ geometries of massive type IIA supergravity. This corresponds to the spectrum of spin-2 operators of the dual Chern-Simons theories. By keeping the geometry (\ref{KKSU3sectorinIIA}) explicitly dependent on the $D=4$ scalar vevs, we will be able to compute a master  graviton mass formula that will depend on those vevs and on the quantum numbers of the generic symmetry group, SU(3). Finding the KK graviton masses for the individual AdS$_4$ solutions \cite{Guarino:2015jca,Varela:2015uca} will then simply entail particularising the master formula to the relevant scalar vevs.

For future reference, let us conclude this section with a discussion of the symmetry properties of the family of metrics (\ref{KKSU3sectorinIIA}), following \cite{Varela:2015uca}. For generic values of the parameters, the metric (\ref{KKSU3sectorinIIA}) displays an $\textrm{SU}(3) \times \textrm{U}(1)$ isometry. The SU(3) factor corresponds to the isometries of the $ds^2 ( \mathbb{CP}^2 ) $ part of the geometry, while the U(1) is generated by the Killing vector $\partial_\psi$. Note that the metric preserves this U(1) in spite of depending on the charged scalars $\zeta$, $\tilde \zeta$, since it only depends on them through the U(1)--invariant combination $\zeta^2 + \tilde\zeta^2$. The IIA fluxes \cite{Varela:2015uca} do generically break this U(1). Symmetry enhancements occur by restricting the parameters to certain submanifolds of (\ref{scalarmanifold}). On the surface 
\begin{eqnarray} \label{eq:SO6configs}
\chi =  \zeta = \tilde \zeta = 0 \; ,
\end{eqnarray}
the metric (\ref{KKSU3sectorinIIA}) reduces to 
{\setlength\arraycolsep{2pt}
\begin{eqnarray} \label{KKSO6sectorinIIA}
d\bar{s}_{6}^2 =  L^{-2} g^{-2} \,   \Big[ \;  e^{-2\phi+\varphi}  \,  d\alpha^2      +  \Delta^{-1}_2   \sin^2 \alpha \,   d\tilde{s}^2 ( S^5 )  \Big] \; , 
\end{eqnarray}
}where $d\tilde{s}^2 ( S^5 )$ is the round Einstein metric on the unit $S^5$. The metric (\ref{KKSO6sectorinIIA}) indeed displays an enhanced SO(6) isometry group which rotates the $S^5$. Finally, for 
\begin{eqnarray} \label{eq:SU3toG2}
\phi = \varphi \; , \qquad 
\tilde \zeta = 2\chi \; , \qquad 
a= \zeta = 0 \; ,
\end{eqnarray}
the geometry (\ref{KKSU3sectorinIIA}) becomes
{\setlength\arraycolsep{2pt}
\begin{eqnarray} \label{KKG2sectorinIIA}
d\bar{s}_{6}^2 =  L^{-2} g^{-2} \,   e^{-\varphi} \big( 1+e^{2 \varphi} \chi^2 \big)^{-1}   ds^2(S^6) \; , 
\end{eqnarray}
}where $ds^2 ( S^6 )$ is the round Einstein metric on the unit $S^6$. The isometry of this configuration is therefore SO(7). The IIA fluxes, however, generically break this to G$_2 \subset \textrm{SO}(7)$, unless $\chi = 0$ is further imposed.

\subsection{Boundary value problem}

On the geometry (\ref{KKSU3sectorinIIA}), the partial differential equation (PDE) (\ref{FPeq}) becomes
\be \label{SU3eqMassGrav}
g^2 \Big(Xe^{2\phi-\varphi} \big( \partial^2_{\alpha}+5\cot\alpha \, \partial_{\alpha} \big)
+\frac{\Delta_1}{\sin^2\alpha} \, \Box_{S^5} +\frac{X\Delta_2-\Delta_1}{\sin^2\alpha} \, \partial^2_{\psi}  \Big)Y(y)=-M^2Y(y)\,,
\ee
as shown in appendix \ref{app:massop}. Here $\Box_{S^5}$ is the scalar Laplacian on the unit radius five-sphere. As we will next argue, this PDE turns out to be separable.

The appearance of the Laplacian $\Box_{S^5}$ in the PDE (\ref{SU3eqMassGrav}) suggests that the eigenfunction $Y(y)$ should be expandable in terms of the $S^5$ spherical harmonics $Y_\ell (\tilde \mu^i )$. These are polynomials of the $\mathbb{R}^6$ coordinates $\tilde \mu^i$, $i=1 , \ldots , 6$, that define $S^5$ via the constraint $\delta_{ij} \tilde{\mu}^i  \tilde{\mu}^j = 1$, and span the symmetric-traceless representation $[0, \ell, 0]$ of $\textrm{SU}(4) \sim \textrm{SO}(6)$. The presence of the operator $\partial^2_{\psi} $, however, generically reduces the symmetry of the problem down to $\textrm{SU}(3) \times \textrm{U}(1) \subset \textrm{SO}(6)$. Accordingly, the eigenfunctions should come in representations of this smaller symmetry group. Thus, we still expect the eigenfunction $Y(y)$ to be expandable in $S^5$ spherical harmonics $Y_\ell (\tilde \mu^i )$, but with eigenspaces split according to
\begin{equation}
  \label{eq:SO6toU3branching}
  [0, \ell, 0] \;
  \stackrel{\mathrm{SU}(3)\times\mathrm{U}(1)}{\longrightarrow} \; 
   \sum_{p=0}^\ell \, [p, \ell -p ]_{\frac{2}{3} (\ell-2p) } \; , 
\end{equation}
where the subscript indicates the U(1) charge, suitably normalised. More concretely, this is the normalisation with respect to the Killing vector $\partial_{\tilde \psi} = -\frac23 \partial_{\psi}$. 

 In order to implement the splitting (\ref{eq:SO6toU3branching}) in practice, we introduce complex coordinates $z^a$ and their conjugates $\bar{z}_a$, $a = 1,2,3$, as $z^1 = \tilde{\mu}^1 + i \tilde{\mu}^2$, etc., and write the  $[0, \ell, 0]$ spherical harmonics on $S^5$ as a polynomial in $z^a$, $\bar{z}_a$, 
\begin{eqnarray} \label{S5SphericalHarm}
Y_{\ell , p}  \, (z,\bar{z} ) &=&  c_{a_1 \cdots a_p}{}^{b_1 \cdots b_{\ell - p}}  \, z^{a_1} \cdots z^{a_p}  \, \bar{z}_{b_1} \cdots \bar{z}_{b_{\ell-p}} \; ,
\end{eqnarray}
where $c_{a_1 \cdots a_p}{}^{b_1 \cdots b_{\ell - p}}$ are constants in the $[q,p]$ representation of SU(3), with $q \equiv \ell-p$, and 
\begin{equation} \label{rangep}
p = 0 , 1 , \ldots , \ell \; ,
\end{equation}
as follows from (\ref{eq:SO6toU3branching}). In the basis (\ref{S5SphericalHarm}), $\textrm{SU}(3) \times \textrm{U}(1)$ acts diagonally, in the sense that $Y_{\ell , p}$ are, of course, eigenfunctions of the $S^5$ Laplacian,
\begin{equation} \label{EigenS5}
\Box_{S^5} \, Y_{\ell , p}(z,\bar{z}) = -\ell (\ell +4) \, Y_{\ell , p } (z,\bar{z}) \; , 
\end{equation}
which also have definite U(1) charge,
\begin{equation} \label{Eigenpsi}
\partial^2_{\psi} \, Y_{\ell , p}(z,\bar{z})  = -(\ell - 2p)^2 \,  Y_{\ell , p }(z,\bar{z}) \; .
\end{equation}

This discussion leads us to consider the following factorised form for the eigenfunction in (\ref{SU3eqMassGrav}): 
\begin{equation} \label{EFansatz1}
Y(\alpha, z,\bar{z} ) = f (\alpha) \, Y_{\ell, p} (z,\bar{z}) \; ,
\end{equation}
where $f(\alpha)$ is a function of the angle $\alpha$, and we have suppressed the $\ell$, $p$ labels on the left-hand-side. Inserting (\ref{EFansatz1}) into (\ref{SU3eqMassGrav}) and making use of the eigenfunction conditions 
(\ref{EigenS5}), (\ref{Eigenpsi}), equation (\ref{SU3eqMassGrav}) becomes an ordinary differential equation (ODE) on $\alpha$:
\begin{equation} \label{SU3eqMassGravSplit}
Xe^{2\phi-\varphi} \big( f^{\prime\prime}(\alpha) +5\cot\alpha \, f^{\prime}(\alpha) \big)
- \Big(\ell (\ell+4 )  \frac{\Delta_1}{\sin^2\alpha} +(\ell - 2p)^2 \frac{X\Delta_2-\Delta_1}{\sin^2\alpha} \Big) f(\alpha) = -g^{-2} M^2f(\alpha) \, .
\end{equation}
We have thus reduced our eigenvalue problem to solving the ODE (\ref{SU3eqMassGravSplit}) with specific boundary conditions: those ensuring regularity of the eigenfunction.

Next, we move on to solve the ODE (\ref{SU3eqMassGravSplit}). In order to do this, we change variables as
\be \label{Changvars}
u=\cos^2\alpha \; ,
\qquad f(\alpha)  = (1-u)^{\frac{\ell}2} \, H(u) \; .
\ee
This change brings (\ref{SU3eqMassGravSplit}) into standard hypergometric form,
\be \label{HyperGeometricODE}
(1-u)u \, H''(u)+(c-(1+a+b)u) \, H'(u) -ab \, H(u)=0,
\ee
where the constants $a$, $b$, $c$ are given in terms of the integers $\ell$ and $p$ and the (\ref{scalarmanifold})--valued scalar vevs by
\begin{equation} \label{HGparams}
a = \tfrac14 (2 \ell +5 ) - \tfrac12 \, e^{ \frac12 \varphi -\phi  } \, X^{-\frac12} \sqrt{ \Xi   } \; , \qquad 
b = \tfrac14 (2 \ell +5 ) + \tfrac12 \, e^{ \frac12 \varphi -\phi  } \, X^{-\frac12} \sqrt{ \Xi   } \; , \qquad 
c = \tfrac12 \; .
\end{equation}
Here we have defined 
\begin{equation} \label{HGparamsDef}
\Xi \equiv  M^2 g^{-2} +\tfrac{25}{4} e^{2\phi-\varphi} X  - \big( e^{\varphi} Y -e^{2\phi-\varphi} X  \big)  \ell (\ell+4)  - e^{\varphi} \big( X- Y  \big)  (\ell - 2p)^2   \; . 
\end{equation}
The two linearly independent solutions to the hypergeometric ODE (\ref{HyperGeometricODE}) are given by the hypergeometric functions
\be \label{LinIndepeHGSols}
 {}_2F_1(a,b,c ; u)\, \quad {\rm and}\quad  u^{1-c}{}_{\, 2} F_1(1+a-c,1+b-c,2-c ; u) \; .
\ee

Finally, we impose boundary conditions to ensure regularity. The relevant range of $u$ is $0 \leq u \leq 1$ (by (\ref{Changvars}), the original coordinate $\alpha$ in (\ref{anglerange}) covers this range twice). Both linearly independent solutions (\ref{LinIndepeHGSols}) are regular\footnote{This is unlike in \cite{Klebanov:2009kp,Ahn:2009et,Pang:2015rwd}, where the second solution is singular at $u=0$ and is thus discarded.} at $u=0$ for all values of the parameters. Regularity at the other end, however, can only be achieved through appropriate restrictions on the parameters. Regularity of the first solution at $u=1$ requires setting $a = - j $, with $j$ a non-negative integer. Imposing this condition in (\ref{HGparams}), (\ref{HGparamsDef}), we find a first tower of generic KK graviton squared masses:
\begin{equation} \label{KKevenbranch}
 g^{-2} M^2_{\1 \, j, \ell , p} =  e^{2\phi-\varphi} X (2j+\ell) (2 j + \ell +5)    + \big( e^{\varphi} Y -e^{2\phi-\varphi} X  \big)  \ell (\ell+4)  + e^{\varphi} \big( X- Y  \big)  (\ell - 2p)^2  \; .
\end{equation} 
The corresponding eigenfunctions are given by (\ref{EFansatz1}), (\ref{Changvars}), with $H(u)$ given by the first choice in (\ref{LinIndepeHGSols}), namely,
\begin{eqnarray} \label{EigenFunceven}
Y_{\1 \, j, \ell , p}  \, (\alpha, z,\bar{z} ) &=&  c_{a_1 \cdots a_p}{}^{b_1 \cdots b_{\ell - p}}  \, z^{a_1} \cdots z^{a_p}  \, \bar{z}_{b_1} \cdots \bar{z}_{b_{\ell-p}}   \nonumber \\
&&  \times   \sin^\ell \alpha \sum_{k=0}^j (-1)^k \, { j \choose k } \, \frac{ \left( j+ \ell +\tfrac52 \right)_k }{ \left( \tfrac12 \right)_k } \, \cos^{2k} \alpha \; ,
\end{eqnarray}
where
\begin{equation} \label{Pochhammer}
(x)_k =
\left\{
\begin{array}{lll}
1 & ,                        & \textrm{if $k = 0$}  \\
x (x+1) \cdots (x+k-1)  & ,
& \textrm{if  $k  > 0$ }
 \end{array} \right.
\end{equation}
is the Pochhammer symbol. Regularity of the second solution in (\ref{LinIndepeHGSols}) at $u=1$ requires $1+a-c = - j $, with $j$ again a non-negative integer. Bringing this condition to (\ref{HGparams}), (\ref{HGparamsDef}), we find a second tower of  generic KK graviton squared masses:
\begin{equation} \label{KKoddbranch}
g^{-2} M^2_{\2 \, j, \ell , p} =  e^{2\phi-\varphi} X (2j+ 1+ \ell) (2 j + \ell +6)   + \big( e^{\varphi} Y -e^{2\phi-\varphi} X  \big)  \ell (\ell+4)  + e^{\varphi} \big( X- Y  \big)  (\ell - 2p)^2  \; .
\end{equation} 
The associated eigenfunctions are given by (\ref{EFansatz1}), (\ref{Changvars}), with $H(u)$ given by the second choice in (\ref{LinIndepeHGSols}):
\begin{eqnarray} \label{EigenFuncodd}
Y_{\2 \, j, \ell , p}  \, (\alpha, z,\bar{z} ) &=&  c_{a_1 \cdots a_p}{}^{b_1 \cdots b_{\ell - p}}  \, z^{a_1} \cdots z^{a_p}  \, \bar{z}_{b_1} \cdots \bar{z}_{b_{\ell-p}}   \nonumber \\
&&  \times   \sin^\ell \alpha \sum_{k=0}^j (-1)^k \, { j \choose k } \, \frac{ \left( j+ \ell +\tfrac72 \right)_k }{ \left( \tfrac32 \right)_k } \, \cos^{2k+1} \alpha \; .
\end{eqnarray}

\subsection{Final form and completeness of the generic solution} \label{sec:FinalFormComplete}

A quick inspection of the eigenvalues (\ref{KKevenbranch}) and (\ref{KKoddbranch}) makes it obvious that these two series in fact correspond to one and only branch of KK graviton masses. Indeed, trading $j$ for a new integer $n$ defined for convenience as 
\begin{equation} \label{nQN}
n =
\left\{
\begin{array}{lll}
2j + \ell  & ,                        & \textrm{for the first branch}  \\
2j + 1 + \ell   & ,                 & \textrm{for the second branch} \; ,
 \end{array} \right.
\end{equation}
(\ref{KKevenbranch}) and (\ref{KKoddbranch}) can be combined into the single KK tower:
\begin{equation} \label{KKbranch}
 g^{-2} M^2_{n, \ell , p} = e^{2\phi-\varphi} X n  (n +5)  + \big( e^{\varphi} Y -e^{2\phi-\varphi} X  \big)  \ell (\ell+4)  + e^{\varphi} \big( X- Y  \big)  (\ell - 2p)^2  \; ,
\end{equation} 
where it is important to note that the quantum numbers range as
\begin{eqnarray} \label{QNranges}
n = 0, 1, 2 , \ldots \; , \qquad
\ell = 0 , 1 , \ldots , n \; , \qquad 
p = 0 , 1 , \ldots , \ell \; .
\end{eqnarray}
Only $n$ ranges freely over the non-negative integers, due to its definition (\ref{nQN}) in terms of the non-negative but otherwise unconstrained integer $j$. The range of $p$ was all along constrained by $\ell$ by equation (\ref{eq:SO6toU3branching}) (see (\ref{rangep})), and the range of $\ell$ turns out to be limited by $n$ since, again by (\ref{nQN}), $n \geq \ell$. At fixed $n$, the eigenvalue (\ref{KKbranch}) occurs with degeneracy
\begin{equation} \label{MassDeg}
d_{\ell,p} \equiv \textrm{dim} \, [p, \ell - p]  = \tfrac12 (p+1) (\ell -p +1) (\ell+2) \; .
\end{equation}

Similarly, the two eigenfunction branches (\ref{EigenFunceven}), (\ref{EigenFuncodd}) can be combined into a single formula. Defining
\begin{eqnarray}
h_{n , \ell} \equiv n - \ell - 2 \left[  \frac{n-\ell}{2}\right] = 
\left\{
\begin{array}{lll}
0  & ,                        & \textrm{if $n- \ell$ is even, as in the first branch} \; ,  \\
1   & ,                 & \textrm{if $n- \ell$ is odd, as in the second branch} \; ,
 \end{array} \right.
\end{eqnarray}
where the square brackets denote integer part, the eigenfunction corresponding to the squared mass (\ref{KKbranch}) can be compactly written as 
\begin{eqnarray} \label{EigenFunc}
Y_{n, \ell , p}  \, (\alpha, z,\bar{z} ) &=&  c_{a_1 \cdots a_p}{}^{b_1 \cdots b_{\ell - p}}  \, z^{a_1} \cdots z^{a_p}  \, \bar{z}_{b_1} \cdots \bar{z}_{b_{\ell-p}}    \\
&&  \times   \sin^\ell \alpha \sum_{k=0}^{ \left[  \frac{n-\ell}{2}\right]  }  (-1)^k \, { \left[  \frac{n-\ell}{2}\right]   \choose k } \, \frac{ \left( \left[  \frac{n-\ell}{2}\right]  + \ell +\tfrac52 + h_{n , \ell}  \right)_k }{ \left( \tfrac12 + h_{n , \ell}  \right)_k } \, \cos^{2k + h_{n , \ell}  } \alpha \; . \nonumber
\end{eqnarray}
For later purposes, it is convenient to present an alternate form for this eigenfunction in terms of (constrained) coordinates on $\mathbb{R}^7$. Let\footnote{Although redundant, we present both notations $X^I$ and $\mu^I$ as both are often used in the literature.} $X^I \equiv \mu^I$, $I=1 , \ldots , 7$, parametrise the directions transverse to the D2-branes, subject to the $S^6$ constraint $\delta_{IJ} \mu^I \mu^J = 1$. These $\mu^I$ can be written in terms of the $\tilde{\mu}^i$ defined above (\ref{eq:SO6toU3branching}) and the angle $\alpha$ as 
\begin{eqnarray} \label{musplit}
X^i \equiv \mu^ i = \sin \alpha \, \tilde{\mu}^i \; , \; i = 1, \ldots, 6 \; ,  \qquad 
X^7 \equiv \mu^7 = \cos \alpha \; .
\end{eqnarray}
The first six directions can be complexified as $Z^a = z^a  \sin \alpha$, $a=1,2,3$, in terms of the $z^a$ written above (\ref{S5SphericalHarm}). In terms of these, the eigenfunction (\ref{EigenFunc}) can be rewritten as
\begin{eqnarray} \label{EigenFuncRewrite}
Y_{n, \ell , p}  \, (Z,\bar{Z} , X^7 ) &=&  c_{a_1 \cdots a_p}{}^{b_1 \cdots b_{\ell - p}}  \, Z^{a_1} \cdots Z^{a_p}  \, \bar{Z}_{b_1} \cdots \bar{Z}_{b_{\ell-p}}    \\
&&  \times  \sum_{k=0}^{ \left[  \frac{n-\ell}{2}\right]  }  (-1)^k \, { \left[  \frac{n-\ell}{2}\right]   \choose k } \, \frac{ \left( \left[  \frac{n-\ell}{2}\right]  + \ell +\tfrac52 + h_{n , \ell}  \right)_k }{ \left( \tfrac12 + h_{n , \ell}  \right)_k } \, (X^7)^{2k + h_{n , \ell}  } \; , \nonumber
\end{eqnarray}
depending implicitly on $(\alpha, z^a ,\bar{z}_a )$ through $(Z^a ,\bar{Z}_a , X^7 )$. 

We must still argue that the solution (\ref{KKbranch}), (\ref{EigenFunc}) to the boundary value problem is complete. We will argue for completeness of the spectrum based on its dependence on the relevant quantum numbers, and on symmetry considerations. The key observation is that, unlike $j$, the quantum number $n$ enjoys a precise interpretation: it corresponds to the Kaluza-Klein level. Namely, $n$ turns out to be the Dynkin label of the symmetric traceless representation $[n,0,0]$ of SO(7), the largest symmetry that can be imposed on our problem. As the KK level, $n$ allows for a systematic arrangement of the spectrum. The easiest way to see this role of $n$ is by particularising the problem to the SO(7)--invariant subspace (\ref{eq:SU3toG2}) of the scalar manifold (\ref{scalarmanifold}). With this restriction, the resulting internal background metric (\ref{KKG2sectorinIIA}) becomes proportional to the SO(7)--invariant metric on the round $S^6$, and the spectrum is given by the SO(7) spherical harmonics. Indeed, under this assumption, the eigenvalue (\ref{KKbranch}) scales solely with the characteristic $n(n+5)$ dependence of the $S^6$ spherical harmonic eigenvalues,
\begin{equation} \label{KKEigenvalueSO7}
 g^{-2} M^2_{n} = e^{\varphi} X n  (n +5)  \; ,
\end{equation} 
and the eigenfunctions (\ref{EigenFunc}) combine into the $S^6$ spherical harmonics, 
\begin{equation} \label{KKEigenfunctionSO7}
Y_n (\mu^I) = c_{I_1 \cdots I_n} \, \mu^{I_1} \cdots \mu^{I_n} \; , 
\end{equation}
with $c_{I_1 \cdots I_n}$ constants in the $[n,0,0]$ representation of SO(7). At fixed KK level $n$, the degeneracy of the SO(7)--symmetric spectrum is 
\begin{equation} \label{DegeneracySO7}
D_{n , 7} \equiv \textrm{dim} \, [n, 0, 0] = { n+6   \choose n } - { n+4   \choose n -2 }  = \tfrac{1}{5!} (2n+5)(n+4)(n+3)(n+2)(n+1) ,
\end{equation}
where, more generally and for future reference, $D_{k,N}$ is the dimension of the symmetric traceless representation $[k,0, \ldots, 0]$ of SO$(N)$,
\begin{eqnarray} \label{kSONdim}
D_{k , N} &=  & { k+N-1   \choose k } - { k+N-3   \choose k -2 }  \\[5pt]
&=& \tfrac{1}{(N-2)!} \, (2k+N-2) (k+N-3) (k+N-4) \cdots (k+2)(k+1) \; . \nonumber
\end{eqnarray}

The completeness of the SO(7)--symmetric spectrum (\ref{KKEigenvalueSO7}), (\ref{KKEigenfunctionSO7}) is apparent. The completeness of the generic spectrum with only $\textrm{SU}(3) \times \textrm{U}(1) \subset \textrm{SO}(7)$ symmetry, (\ref{KKbranch}), (\ref{EigenFunc}), also follows. The generic spectrum comes in the representations of $\textrm{SU}(3) \times \textrm{U}(1)$ that result from branching the symmetric traceless representation $[n,0,0]$ of SO(7) for each $n$ through $\textrm{SO}(6) \sim \textrm{SU} (4) $ and then through (\ref{eq:SO6toU3branching}), that is,
\begin{equation}
  \label{eq:SO7toSO6toU3branching}
  [n, 0, 0] \;
  \stackrel{\mathrm{SU}(4)}{\longrightarrow} \; 
   \sum_{\ell=0}^n \, [0, \ell ,0] \; 
  \stackrel{\mathrm{SU}(3)\times\mathrm{U}(1)}{\longrightarrow} \; 
   \sum_{\ell=0}^n  \sum_{p=0}^\ell \, [p, \ell -p ]_{\frac23(\ell - 2p)} \; .
\end{equation}
This follows from the most general expression, (\ref{KKEmbeddingFulleqn}), that the mass operator of this class of geometries may have. This is also consistent with the quantum number ranges (\ref{QNranges}). Accordingly, the generic degeneracies (\ref{MassDeg}) are related to the degeneracy (\ref{DegeneracySO7}) of the SO(7)--symmetric spectrum as
\begin{equation}
D_{n,7} =  \sum_{\ell=0}^n \sum_{p=0}^\ell d_{\ell , p} \; .
\end{equation}
No other $\textrm{SU}(3) \times \textrm{U}(1)$ state arises that cannot be tracked down to descend from a symmetric traceless representation of $[n,0,0]$ for some $n$ via (\ref{eq:SO7toSO6toU3branching}). Finally, the eigenfunctions (\ref{EigenFunc}) are the $S^6$ spherical harmonics (\ref{KKEigenfunctionSO7}), branched out into $\textrm{SU}(3) \times \textrm{U}(1)$ representations via (\ref{eq:SO7toSO6toU3branching}) through the split (\ref{musplit}) and the identifications in terms of $z^a$, $\bar{z}_a$ written above (\ref{S5SphericalHarm}). In particular, the eigenfunctions (\ref{EigenFunc}) are polynomials in $z^a$, $\bar{z}_a$, $\sin \alpha$, $\cos \alpha$.

\subsection{Summary} \label{sec:summary}

To summarise, the complete spectrum of transverse, traceless gravitons on the background AdS$_4$ geometries (\ref{subsec:Setup}), (\ref{KKSU3sectorinIIA}) of massive type IIA supergravity is given by the KK tower
\begin{eqnarray} \label{KKTowerExpansion}
h_{\mu\nu} (x, \alpha, z,\bar{z} ) = \sum_{n=0}^\infty \sum_{\ell=0}^n \sum_{p=0}^\ell \,  h_{\mu\nu \, n, \ell , p }^{[\textrm{tt}]} (x) \, Y_{n, \ell , p}  \, (\alpha, z,\bar{z} ) \; , 
\end{eqnarray}
where the complete set of eigenfunctions $Y_{n, \ell , p}  \, (\alpha, z,\bar{z} ) $ is defined in (\ref{EigenFunc}). These  
correspond to the $S^6$ spherical harmonics branched out into $\textrm{SU}(3) \times \textrm{U}(1)$ representations through (\ref{eq:SO7toSO6toU3branching}), with degeneracies $d_{\ell,p}$ given in (\ref{MassDeg}). The corresponding graviton squared masses are the $M^2_{n, \ell , p}$ written in (\ref{KKbranch}). The generic spectrum depends non-linearly on the (\ref{scalarmanifold})--valued vevs of the $\textrm{SU}(3) \times \textrm{U}(1)$-invariant scalars of $D=4$ $\cN=8$ dyonically-gauged ISO$(7)$ supergravity, and quadratically on three quantum numbers $n$, $\ell$, $p$ with ranges (\ref{QNranges}). The integer $n$ is the KK level, {\it i.e.}, it is the Dynkin label of the symmetric traceless representation of the maximal symmetry group SO(7). The integer $\ell$ is the Dynkin label of the symmetric traceless representation of SO(6), and $p$ labels the $\textrm{SU}(3)$ representations. The $\textrm{U}(1)$ charge is not an independent quantum number, it is fixed by $\ell$ and $p$ as in (\ref{eq:SO7toSO6toU3branching}).

On the surface (\ref{eq:SO6configs}) of the parameter space (\ref{scalarmanifold}), the symmetry of the problem is enhanced to SO(6). Accordingly, at each KK level $n$, the spectrum is organised in SO(6) representations via the first branching in (\ref{eq:SO7toSO6toU3branching}). The term in $(\ell - 2p)^2$ coming from the U(1) charge drops out from the eigenvalue (\ref{KKbranch}), and only the terms in $n(n+5)$ and $\ell ( \ell + 4)$ remain. Similarly, the eigenfunctions (\ref{EigenFunc}) simply combine into $\mu^{ \{ I_1} \cdots \mu^{ I_n \} }$ with the $\mu^I$ split as in (\ref{musplit}). If the symmetry is further enhaced to SO(7) by imposing the restrictions (\ref{eq:SU3toG2}), then the term in $\ell ( \ell + 4)$ also drops out from the eigenvalue (\ref{KKbranch}), and the only remaining term is that in $n(n+5)$, see (\ref{KKEigenvalueSO7}). The eigenfunctions in the latter case become the SO(7)--irreducible spherical harmonics (\ref{KKEigenfunctionSO7}) on the round $S^6$.


\section{Graviton mass spectrum of individual solutions} \label{sec:IndivSols}

Having worked out the generic problem, we now turn to obtaining the specific KK graviton spectrum for each of the  AdS$_4$ solutions of massive IIA supergravity that uplift from vacua of $D=4$ $\cN=8$ dyonically-gauged ISO(7) supergravity with at least SU(3) symmetry.

\subsection{KK graviton masses}

Recall from \cite{Guarino:2015qaa} that the SU(3)--invariant sector of the $\cN=8$ supergravity contains critical points with residual supersymmetry and bosonic symmetry $(\cN=2 , \textrm{SU}(3) \times \textrm{U}(1))$,  $(\cN=1 , \textrm{G}_2)$ and $(\cN=1 , \textrm{SU}(3))$. In addition, it also contains non-supersymmetric critical points with residual symmetry\footnote{The first two were denoted SO(7)$_+$, $\textrm{SO}(6)_+$ in \cite{Guarino:2015qaa}. Here we change the notation following appendix \ref{sec:SO8triality}. For similar reasons, the U(1) factor of the $\cN=2$ solution could be denoted as U$(1)_v$, but we drop the label $v$ in this case.} SO(7)$_v$, $\textrm{SO}(6)_v$, G$_2$ and SU(3), the latter only known numerically. These solutions were uplifted \cite{Guarino:2015jca,Varela:2015uca} using the consistent truncation of \cite{Guarino:2015jca,Guarino:2015vca} to obtain new AdS$_4$ solutions \cite{Guarino:2015jca,Varela:2015uca} of massive type IIA supergravity and recover previously known ones \cite{Romans:1985tz,Behrndt:2004km,Lust:2008zd}. The ten-dimensional solutions are obtained by evaluating the explicit SU(3)--invariant consistent uplift formulae of \cite{Varela:2015uca} at the corresponding vevs of the $D=4$ scalars, recorded in table 3 of \cite{Guarino:2015qaa}. Similarly, we can evaluate the master formula (\ref{KKbranch}) on the $D=4$ scalar vevs for each solution to obtain its spectrum of gravitons. The result, including the analytically known non-supersymmetric solutions for completeness, is
\begin{equation} \label{eq:KKGravSpectraSols}
\textrm{
\begin{tabular}{llll}
$\cN=2 \ , \; \textrm{SU}(3) \times \textrm{U}(1) $  & : &  
$
L^2 M_{n,\ell,p}^2 = \tfrac23  n  ( n + 5) -\tfrac13    \ell  ( \ell + 4)  + \tfrac19  (\ell - 2p)^2
$  & ,  
  \\[10pt]
$\cN=1 \ , \; \textrm{G}_2$  & : &  
$
L^2 M_{n}^2 = \tfrac{5}{12}  n  ( n + 5) 
$  & ,  \\[10pt]
$\cN=1 \ , \; \textrm{SU}(3)$  & : &  
$
L^2 M_{n,\ell,p}^2 = \tfrac56  n  ( n + 5) -\tfrac{5}{12}    \ell  ( \ell + 4)  - \tfrac{5}{36}  (\ell - 2p)^2
$  & , 
 \\[20pt]
$\cN=0 \ , \; \textrm{SO}(7)_v$  & : &  
$
L^2 M_{n}^2 = \tfrac{2}{5}  n  ( n + 5)
$  & , 
\\[10pt]
$\cN=0 \ , \; \textrm{SO}(6)_v$  & : &  
$
L^2 M_{n,\ell}^2 =  n  ( n + 5) -\tfrac{3}{4}    \ell  ( \ell + 4)  
$  & , 
\\[10pt]
$\cN=0 \ , \; \textrm{G}_2$  & : &  
$
L^2 M_{n}^2 = \tfrac{1}{2}  n  ( n + 5) 
$  & , 
%
%
\end{tabular}
}
\end{equation}
with the quantum numbers ranging as in (\ref{QNranges}). All these graviton spectra are new. Note that, for solutions with enhanced symmetry, the dependence on some of the quantum numbers drops out following the pattern discussed in section \ref{sec:summary}. The conformal dimensions $\Delta$ of the corresponding dual operators are given by the largest root of the equation
\begin{eqnarray} \label{DeltaM}
\Delta (\Delta - 3) = M^2 L^2 \; , 
\end{eqnarray}
where $M^2$ denotes each of the eigenvalues in (\ref{eq:KKGravSpectraSols}). Finally, note that at KK level $n=0$ all solutions display, as expected, a massless graviton which is a singlet of the residual symmetry group.

\subsection{$\cN=2$ spin-two spectrum and dual operators} \label{N=2spin2sectrum}

The solutions that preserve some supersymmetry $\cN$ and residual bosonic symmetry $G$ must have their spectrum fall in irreducible representations of $\textrm{OSp}(4|\cN) \times G$ including states of at most spin 2. Recall that, for $\textrm{OSp}(4|1)$, a massless graviton partners with a massless gravitino, and a massive graviton of energy  $\Delta$ partners with two gravitini of energies $\Delta \pm \frac12 $ and a vector of energy $\Delta$, see {\it e.g.} \cite{Duff:1986hr}. At given KK level $n$, these multiplets have the $\Delta$ that follows from (\ref{eq:KKGravSpectraSols}) via (\ref{DeltaM}), and occur in the $[n,0]$ irrep of G$_2$, for the $\cN=1 , \textrm{G}_2$ solution, and in the $[p,\ell-p]$, $\ell = 0 , 1 , \ldots , n$, $p = 0 , 1 , \ldots , \ell$, irrep of SU(3) for the $\cN=1 , \textrm{SU}(3)$ solution.

More interesting is the situation for the $\cN=2$, $\textrm{SU}(3) \times \textrm{U}(1)$ solution. Shortening can occur in this case, leading to three possible types of $\textrm{OSp}(4|2)$ supermultiplets containing states of up to spin-2: a massless graviton multiplet, and short and long massive graviton multiplets. See \cite{Freedman:1983na,Ceresole:1984hr,Nicolai:1985hs} for the Osp$(4|2)$ representation theory and appendix A of \cite{Klebanov:2008vq} for a convenient summary. At KK level $n=0$, the massless, $\Delta=3$, graviton partners with two massless gravitini and a massless vector into a massless graviton multiplet, see table 8 of \cite{Klebanov:2008vq}. At higher KK levels, massive gravitons lie into either short or long multiplets, depending on whether or not their energy $\Delta$ and U(1) R-charge $R$ saturate the bound $\Delta \geq |R| + 3$. See tables 9 and 10 of \cite{Klebanov:2008vq} for the field content of these multiplets.

 \begin{table}[]
\centering

\resizebox{\textwidth}{!}{

\begin{tabular}{|c|l|c|c|c|c|c|c|}
\hline
$n$                & $[p,\ell-p]_{\frac{2}{3}(\ell-2p)}$                  & $d_{\ell , p}$ & $L^2M^2_{n,\ell,p}$        & $\Delta_{n,\ell,p}$                                          & $L^2$tr$M^2_{(n)}$                & Dual operator                                                          & Short?       \\ \hline \hline
0                  & $ [0,0]_0$                                  & 1   & 0               & 3                                                 & 0                                 & $\mathcal{T}^{(0)}_{\alpha\beta}\vert_{s=2}$                      & $\checkmark$ \\ \hline
\multirow{2}{*}{1} & $ [0,0]_0$                                  & 1   & 4               & 4                                                 & \multirow{2}{*}{$\frac{56}{3}$}   &                                 $T_{\mu\nu}X^7$                                  &              \\ \cline{2-5} \cline{7-8} 
                   & $ [1,0]_{-\frac{2}{3}},[0,1]_{\frac{2}{3}}$ & 3   & $\frac{22}{9}$  & $\frac{11}{3}$                                    &                                   & $\mathcal{T}^{(0)}_{\alpha\beta}\mathcal{Z}^a\vert_{s=2}$, c.c.             & $\checkmark$ \\ \hline
\multirow{4}{*}{2} & $ [0,0]_0$                                  & 1   & $\frac{28}{3}$  & $\frac{1}{2} \left(\sqrt{\frac{139}{3}}+3\right)$ & \multirow{4}{*}{168}              &             $T_{\mu\nu}(Z^a\bar{Z}_a-6(X^7)^2)$                                                      &              \\ \cline{2-5} \cline{7-8} 
                   & $ [1,0]_{-\frac{2}{3}}, [0,1]_{\frac{2}{3}}$ & 3   & $\frac{70}{9}$  & $\frac{14}{3}$                                    &                                   &                             $T_{\mu\nu}Z^aX^7$, c.c.                                      &              \\ \cline{2-5} \cline{7-8} 
                   & $ [2,0]_{-\frac{4}{3}},[0,2]_{\frac{4}{3}}$ & 6   & $\frac{52}{9}$  & $\frac{13}{3}$                                    &                                   & $\mathcal{T}^{(0)}_{\alpha\beta}\mathcal{Z}^{(a}\mathcal{Z}^{b)}\vert_{s=2}$, c.c.    & $\checkmark$ \\ \cline{2-5} \cline{7-8} 
                   & $ [1,1]_0$                                  & 8   & $\frac{16}{3}$  & $\frac{1}{2} \left(\sqrt{\frac{91}{3}}+3\right)$  &                                   &                              $T_{\mu\nu}(Z^a\bar{Z}_b-\frac{1}{3}\delta^a_{b}Z^c\bar{Z}_c)$                                     &              \\ \hline
\multirow{6}{*}{3} & $ [0,0]_0$                                  & 1   & 16              & $\frac{1}{2} \left(\sqrt{73}+3\right)$            & \multirow{6}{*}{$\frac{2464}{3}$} &                    $T_{\mu\nu}(Z^a\bar{Z}_a-2(X^7)^2)X^7$                                               &              \\ \cline{2-5} \cline{7-8} 
                   & $[1,0]_{-\frac{2}{3}}, [0,1]_{\frac{2}{3}}$ & 3   & $\frac{130}{9}$ & $\frac{1}{2} \left(\frac{\sqrt{601}}{3}+3\right)$ &                                   &                                       $T_{\mu\nu}(Z^b\bar{Z}_b-8(X^7)^2)Z^a$, c.c.                            &              \\ \cline{2-5} \cline{7-8} 
                   & $[2,0]_{-\frac{4}{3}}, [0,2]_{\frac{4}{3}}$ & 6   & $\frac{112}{9}$ & $\frac{16}{3}$                                    &                                   &       $T_{\mu\nu}Z^aZ^bX^7$, c.c.                                                           &              \\ \cline{2-5} \cline{7-8} 
                   & $ [1,1]_0$                                  & 8   & 12              & $\frac{1}{2} \left(\sqrt{57}+3\right)$            &                                   &                       $T_{\mu\nu}(Z^a\bar{Z}_b-\frac{1}{3}\delta^a_{b}Z^c\bar{Z}_c)X^7$                                            &              \\ \cline{2-5} \cline{7-8} 
                   & $ [3,0]_{-2},[0,3]_{2}$                       & 10  & 10              & 5                                                 &                                   & $\mathcal{T}^{(0)}_{\alpha\beta}\mathcal{Z}^{(a}\mathcal{Z}^b\mathcal{Z}^{c)}|_{s=2}$, c.c. & $\checkmark$ \\ \cline{2-5} \cline{7-8} 
                   & $ [2,1]_{-\frac{2}{3}}, [1,2]_{\frac{2}{3}}$ & 15  & $\frac{82}{9}$  & $\frac{1}{2} \left(\frac{\sqrt{409}}{3}+3\right)$ &                                   &                $T_{\mu\nu}(Z^aZ^b\bar{Z_c}-\text{trace})$                                                &              \\ \hline
\end{tabular}

}

\caption{\footnotesize{The spectrum of KK gravitons on the $\cN=2$ $\textrm{SU}(3) \times \textrm{U}(1)$--invariant solution \cite{Guarino:2015jca} of massive type IIA up to KK level $n=3$. For each state, its $\textrm{SU}(3) \times \textrm{U}(1)$ charges (\ref{eq:SO7toSO6toU3branching}), degeneracy (\ref{MassDeg}), mass (\ref{eq:KKGravSpectraSols}) and dimension computed from (\ref{DeltaM}) is given. The trace of the mass matrix (\ref{eq:TrSols}) at level $n$ is also given, and the schematic form of the dual  single-trace spin-2 operators. Checked (unchecked) states belong to short (long) graviton supermultiplets of OSp$(4|2)$.}\normalsize}
\label{tab:KKGravMassSU3U1N=2ISO7}
\end{table}

We have tabulated the KK graviton masses for the $\cN=2$, $\textrm{SU}(3) \times \textrm{U}(1)$--invariant solution up to KK level $n = 3$ in table \ref{tab:KKGravMassSU3U1N=2ISO7}. The U(1) factor corresponds to the R-symmetry. From (\ref{eq:SO6toU3branching}), we see that the charge $R$ under this U(1) is not an independent quantum number, it is rather fixed by $\ell$ and $p$ as $R = \tfrac{2}{3} (\ell -2p)$. From table \ref{tab:KKGravMassSU3U1N=2ISO7} we see that, for any state, its dimension $\Delta$, computed from the mass in (\ref{eq:KKGravSpectraSols}) via (\ref{DeltaM}), indeed satisfies the bound $\Delta \geq |R| + 3$. At each KK level $n$, this bound is saturated whenever  $\ell$ and $p$ take values either ($\ell =n$, $p=0$) or ($\ell = n$, $p=n$). Thus, massive gravitons that fall in short multiplets have $\textrm{SU}(3)\times \textrm{U}(1)$ charges, masses $L^2 M_n^2$ and dimensions $\Delta_n$
\begin{equation} \label{SU3U1chargesConDim}
[n, 0]_{-\frac{2n}{3}} \quad \textrm{or} \quad 
[0, n]_{\frac{2n}{3}} \; ,  \qquad L^2 M_n^2 = \tfrac{2}{9} n (2n+9)  \; , \qquad 
\Delta_n = \tfrac{2}{3} n +3 \; ,
\end{equation}
for $n=1, 2, \ldots$ This series also incorporates naturally the $n=0$ massless graviton multiplet with charge $[0,0]_0$ and dimension $\Delta_0 = 3$, and provides a massive counterpart for it at higher KK levels. For these short multiplets, the conformal dimensions $\Delta_n = \tfrac{2}{3} n +3$ are fixed and protected by the R-charge $R_n = \pm \frac23 n$ as $\Delta_n = |R_n| +3$. All other massive gravitons belong to long multiplets. Their classical dimension is unprotected and indeed renormalised for most of these, as is apparent from the table. Note however the existence for $n=1,2, \ldots$ of a series of long multiplets with ($\ell =n-1$, $p=0$) or ($\ell = n-1$, $p=n-1$) such that $\Delta_n= \frac23(n+5)$ and $R_n = \pm \frac23(n-1)$, so $\Delta_n = |R_n| +4$ and thus seemingly protected. Analogue series of long graviton multiplets with seemingly protected dimensions also appear \cite{Klebanov:2009kp} in the spectrum of the CPW solution \cite{Corrado:2001nv}. 

Selecting ($\ell =n$, $p=0$) and ($\ell = n$, $p=n$) in (\ref{EigenFuncRewrite}), the eigenfunctions corresponding to the gravitons that belong to short multiplets can be seen to be
\begin{eqnarray} \label{EigenFuncShort}
Y_{n} (Z)  =  c_{a_1 \cdots a_n}  \, Z^{a_1} \cdots Z^{a_n}  \; , \qquad
\overline{Y}_n (Z) = c^{a_1 \cdots a_n}  \,  \bar{Z}_{a_1} \cdots \bar{Z}_{a_n} \; ,
\end{eqnarray}
indeed compatible with the SU(3) representation assignments in (\ref{SU3U1chargesConDim}). Also, since the U(1) R-charge of $Y_{n} (Z)$ must be $R(Y_{n} (Z)) = -\frac{2n}{3}$ according to (\ref{SU3U1chargesConDim}), we must have $R(Z^a) = -\frac{2}{3}$. Now, the coordinates $Z^a$ correspond holographically to the lowest components of the chiral superfields $\cZ^a$ of the infrared field theory \cite{Guarino:2015jca}, while the seventh coordinate $X^7$ transverse to the D2-branes belongs to a vector multiplet which is integrated out at low energies. The R-charge assignment $R(\cZ^a) = -\frac{2}{3}$ for the superfield $\cZ^a$, inherited from $Z^a$, is compatible with the requirement that the cubic superpotential of the dual field theory \cite{Guarino:2015jca},
\begin{eqnarray} \label{superPot}
\cW \sim \epsilon_{abc} \, \tr \,  \cZ^a [ \cZ^b  , \cZ^c ]  \; ,
\end{eqnarray}
has R-charge\footnote{R-charge sign conventions are immaterial. In \cite{Guarino:2015jca}, the opposite sign for $R(\cZ^a)$ was chosen.} $\pm2$. This match provides a consistency check of our results. For reasons to be justified very shortly, it is natural to assume that $\cZ^a$ has protected conformal dimension
\begin{eqnarray} \label{DeltaandRofZ}
\Delta ( \cZ^a) = - R  ( \cZ^a)   = \tfrac23 \; , \quad a =1,2,3.
\end{eqnarray}

We are now in a position to discuss the series $n=0,1,2, \ldots$ of spin-2 field theory operators dual to the short graviton multiplets (\ref{SU3U1chargesConDim}), (\ref{EigenFuncShort}). The massless, $n=0$, graviton supermultiplet is of course dual to the stress-energy superfield
\be \label{SETsuperfield}
  \mathcal{T}^{(0)}_{\alpha\beta} = \tr \bar{D}_{(\alpha} \bar{\cZ}_a D_{\beta)} \cZ^a + i \tr \bar{\cZ}_a \!\stackrel{\leftrightarrow}{\partial}_{\alpha\beta}\! \cZ^a \; ,
\ee
an SU(3) singlet with R-charge $R_0= 0$ and protected dimension $\Delta_0= 3$. For higher $n$, the dual superfields can be inferred from the eigenfunctions (\ref{EigenFuncShort}) to be of the form 
\be \label{Tmunun}
\mathcal{T}^{(n) a_1 \cdots a_n}_{\alpha\beta} \sim \tr \, \mathcal{T}^{(0)}_{\alpha\beta}  \, \mathcal{Z}^{ ( a_1} \cdots  \mathcal{Z}^{a_n ) } \; , \quad n = 1,2,3 , \ldots 
\ee
together with the complex conjugates. This series of operators has the $\textrm{SU}(3)\times \textrm{U}(1)$ charges given in (\ref{SU3U1chargesConDim}). It also has the dimension in (\ref{SU3U1chargesConDim}) if $\cZ^a$ is assigned the dimension in (\ref{DeltaandRofZ}). This justifies that choice. This series of spin-2 operators has been summarised in table \ref{tab:KKGravMassSU3U1N=2ISO7}. In the table, $\mathcal{T}_{\alpha\beta}|_{s=2}$ indicates spin-2 component of the stress-energy superfield. Note that, $T_{\mu\nu}$ represent instead the stress-energy operator. For completeness, the table also includes operators in long multiplets, whose form is similarly inferred from the eigenfunction (\ref{EigenFuncRewrite}). Everywhere $X^7$ appears, this symbol is understood to stand for the relevant function of the infrared $\cZ^a$, $\bar{\cZ}^a$ into which the $\cN=8$ super-Yang-Mills scalar $X^7$ is integrated out.

We conclude this section with a comparison to the spectrum of short spin-2 superfields \cite{Klebanov:2009kp,Klebanov:2008vq} of the $\cN=2$ SU(3)--flavoured field theory \cite{Benna:2008zy} dual to the $D=11$ CPW solution \cite{Corrado:2001nv}. Recall that this is a quiver-type $\cN=2$ Chern-Simons theory coupled, like \cite{Guarino:2015jca}, to an SU(3) triplet of chiral superfields $\cZ^a$ which in this case have dimensions and R-charges \cite{Benna:2008zy}
\begin{eqnarray} \label{DeltaandRofZM2}
\Delta ( \cZ^a) = R  ( \cZ^a)   = \tfrac13 \; , \quad a =1,2,3.
\end{eqnarray}
The theory has a sextic superpotential, which indeed has R-charge $2$ with the assignments (\ref{DeltaandRofZM2}). This M2-brane $\cN=2$ field theory \cite{Benna:2008zy} has, like its D2-brane counterpart \cite{Guarino:2015jca}, a series of short spin-2 superfields
with $\textrm{SU}(3)\times \textrm{U}(1)$ charges, dimensions $\Delta_n$ and, for completeness, masses $L^2 M_n^2$ of the corresponding KK gravitons given by \cite{Klebanov:2009kp,Klebanov:2008vq},
\begin{equation} \label{SU3U1chargesConDimM2}
[0, 0]_{n} \quad \textrm{or} \quad 
[0, 0]_{-n} \; ,  \qquad L^2 M_n^2 = n (n + 3)  \; , \qquad 
\Delta_n = n + 3 \; ,
\end{equation}
for $n=0,1,2, \ldots$ For $n=0$, this series contains the energy-momentum superfield, whose expression is identical to that, (\ref{SETsuperfield}), of the $\cN=2$ D2-brane theory. However, for $n \geq 1$, the assignments (\ref{SU3U1chargesConDimM2}) of these short operators as well as the short operators themselves \cite{Klebanov:2009kp,Klebanov:2008vq},
\be \label{TmununM2}
\mathcal{T}^{(n) a_1 \cdots a_n}_{\alpha\beta} \sim \tr \, \mathcal{T}^{(0)}_{\alpha\beta}  \, \big( \epsilon_{abc} \cZ^a \cZ^b \cZ^c  \big)^n  \; , \quad n = 1,2,3 , \ldots 
\ee
are completely different to their D2-brane counterparts (\ref{SU3U1chargesConDim}), (\ref{Tmunun}). Here and throughout we are ignoring contributions from monopole operators, see \cite{Klebanov:2008vq} for a discussion.

Due to the universality properties discussed in the introduction, the $n=0$ KK level mass spectrum for all supergravity fields of spin $s \leq 2$ on the $\cN=2$ AdS$_4$ solutions of massive IIA \cite{Guarino:2015jca} and CPW \cite{Corrado:2001nv} in $D=11$ agree. We have just shown that this universality is lost at higher KK levels: these two solutions have a completely different spectrum of dual spin-2 operators, as expected.


\section{Universality of graviton mass matrix traces} \label{sec:GravMassMat}


We have just seen at the level of the (short) spin-2 spectra that the universality of the $n=0$ KK mass spectrum on the AdS$_4$ solutions of \cite{Corrado:2001nv} and \cite{Guarino:2015jca} is resolved at higher KK levels. We will now see, also at the level of the KK graviton spectra, that a softer form of universality is nevertheless still maintained. Prior to this, we will show that a related type of universality, certainly not apparent, is already present in the KK graviton masses (\ref{eq:KKGravSpectraSols}) for the individual AdS$_4$ solutions of massive IIA supergravity in the class considered.

\subsection{D2--brane cases} \label{sec:GravMassMatD2}

For this discussion, we need to introduce the infinite-dimensional KK graviton mass matrix, $\cM^2$. As argued in section \ref{sec:FinalFormComplete}, the completeness of the spectra is guaranteed by the fact that, at fixed level $n$, the eigenfunctions (\ref{EigenFunc}) branch out from the $S^6$ spherical harmonics (\ref{KKEigenfunctionSO7}) through the splitting (\ref{musplit}). This means that the full, infinite dimensional KK graviton mass matrix $\cM^2$ takes on a block diagonal form KK level by KK level,
\begin{eqnarray} \label{Infinitemassmat}
\cM^2 = \textrm{diag} \Big( M_\0^2 \, , M_\1^2 \,  , \ldots  \, , M_\n^2  \, , \ldots \Big) \; .
\end{eqnarray}
Here, $M_\0^2 = 0 $ corresponds to the massless singlet graviton at the bottom of the KK tower, and $M_\n^2$ is a squared matrix of size $D_{n,7} \times D_{n,7}$, with $D_{n,7}$ the dimension (\ref{DegeneracySO7}) of the symmetric traceless representation $[n,0,0]$ of SO(7). On the surface (\ref{eq:SU3toG2}), where the symmetry is enhanced to SO(7), each block is proportional to the identity matrix of dimension  $D_{n,7}$,
\begin{eqnarray} \label{SO7massmatblock}
g^{-2} \, ( M_\n^2 )_{I_1 \cdots I_n}{}^{J_1 \cdots J_n} = e^\varphi X n (n+5) \, \delta_{ \{ I_1 \cdots I_n \} }^{J_1 \cdots J_n} \; , 
\end{eqnarray}
with eigenvalues (\ref{KKEigenvalueSO7}) and eigenfunctions given by the $S^6$ spherical harmonics (\ref{KKEigenfunctionSO7}). In (\ref{SO7massmatblock}), the curly brackets denote traceless symmetrisation as usual. For the generic SU(3)--symmetric problem, each block $M_\n^2 $ will have a more complicated, non-diagonal form; in any case, its eigenvalues and eigenfunctions are given by (\ref{KKbranch}) and (\ref{EigenFunc}) for $p=0,1, \ldots, \ell$ and $\ell = 0, 1 , \ldots, n$, with $n$ fixed. 

We are now in a position to discuss a curious universality property of the KK graviton mass matrix (\ref{Infinitemassmat}) for the massive IIA, D2-brane solutions at hand: the trace of each block $M_\n^2$ turns out to be proportional to a universal polynomial in $n$, which only differs for each solution in an overall function of the $D=4$ $\cN=8$ supergravity scalar vevs. Let us first discuss the particular case with symmetry enhanced to SO(7). With the restrictions (\ref{eq:SU3toG2}) we immediately obtain, from (\ref{SO7massmatblock}), 
\begin{equation} \label{trMnSO7}
g^{-2} \; \tr \, M_\n^2  = e^\varphi X n (n+5) \cdot \tfrac{1}{5!} (2n+5)(n+4)(n+3)(n+2)(n+1) \; , 
\end{equation}
where we have used (\ref{DegeneracySO7}). This result is straightforward because, according to (\ref{SO7massmatblock}), each block $M^2_\n$ is in this case proportional to the identity matrix of dimension (\ref{DegeneracySO7}) with proportionality coefficient (\ref{KKEigenvalueSO7}). Now, it turns out that the eigenvalue, $n(n+5)$, and degeneracy,  $D_{n,7}$, contributions conspire in such a way that (\ref{trMnSO7}) can be rewritten as 
\begin{equation} \label{trMnSO7bis}
g^{-2} \; \tr \, M_\n^2  = 42\, e^\varphi X \, D_{n-1, \, 9} \; .
\end{equation}
This can be seen by using (\ref{kSONdim}) with $k=n-1$ and $N=9$. 

We do not have an argument as to why $\tr \, M_\n^2$ should be proportional to the dimension of the symmetric traceless representation $[n-1,0,0,0]$ of SO$(9)$. We simply take the notation $D_{n-1, \, 9}$ to be shorthand for the polynomial that appears in (\ref{trMnSO7bis}), which turns out to be given by (\ref{kSONdim}) with $k=n-1$ and $N=9$. More surprisingly, similar conspiracies occur for the general SU(3)--symmetric case at hand, even though $M^2_\n$ is not diagonal any more. We can compute the trace of the block $M^2_\n$ at fixed KK level $n$ even if we do not know its generic expression. The only ingredients we need for this calculation are its eigenvalues $M^2_{n, \ell , p}$, given in (\ref{KKbranch}), and their degeneracy $d_{\ell , p}$, given in (\ref{MassDeg}). We compute
\begin{eqnarray} \label{trMngeneral}
g^{-2} \; \tr \, M^2_\n = g^{-2} \; \sum_{\ell=0}^n \sum_{p=0}^\ell \, M^2_{n, \ell , p} \, d_{ \ell , p } = 6 \big( 2 e^{2\phi - \varphi} X + e^\varphi X +4 e^\varphi Y  \big)  \, D_{n-1, \, 9} \; .
\end{eqnarray}
Here, we have again used (\ref{kSONdim}) with $k=n-1$ and $N=9$ in order to write the result in this compact form. Of course, summing over the appropriate ranges (\ref{QNranges}) for the quantum numbers $\ell$ and $p$ at fixed $n$ is crucial to obtain the result (\ref{trMngeneral}). 

When the restrictions (\ref{eq:SU3toG2}) are imposed, (\ref{trMngeneral}) reduces to the SO(7)--symmetric result (\ref{trMnSO7bis}). One can also evaluate (\ref{trMngeneral}) at the specific critical points of $D=4$ $\cN=8$ ISO(7) supergravity recorded in table 3 of \cite{Guarino:2015qaa}, in order to obtain the trace of the KK graviton mass matrix blocks $M_\n$ at fixed KK level $n$, for each of the corresponding AdS$_4$ solutions of massive type IIA supergravity. We obtain
\begin{equation} \label{eq:TrSols}
\textrm{
\begin{tabular}{llll}
$\cN=2 \ , \; \textrm{SU}(3) \times \textrm{U}(1) $  & : &  
$
L^2 \;  \tr \, M_\n^2 = \tfrac{56}{3}  \, D_{n-1, \, 9}
$  & ,  
  \\[10pt]
$\cN=1 \ , \; \textrm{G}_2$  & : &  
$
L^2 \;  \tr \, M_\n^2 = \tfrac{35}{2}  \, D_{n-1, \, 9}
$  & ,  \\[10pt]
$\cN=1 \ , \; \textrm{SU}(3)$  & : &  
$
L^2 \;  \tr \, M_\n^2 = \tfrac{65}{3}  \, D_{n-1, \, 9}
$  & , 
 \\[20pt]
$\cN=0 \ , \; \textrm{SO}(7)_v$  & : &  
$
L^2 \;  \tr \, M_\n^2 = \tfrac{84}{5}  \, D_{n-1, \, 9}
$  & , 
\\[10pt]
$\cN=0 \ , \; \textrm{SO}(6)_v$  & : &  
$
L^2 \;  \tr \, M_\n^2 = \tfrac{39}{2}  \, D_{n-1, \, 9} 
$  & , 
\\[10pt]
$\cN=0 \ , \; \textrm{G}_2$  & : &  
$
L^2 \;  \tr \, M_\n^2 = 21 \, D_{n-1, \, 9}
$  & .
%
%
\end{tabular}
}
\end{equation}
For all the solutions under consideration, the trace of the graviton mass matrix at fixed KK level $n$ turns out to be given by a (7th order) universal polynomial in $n$ which only differs for each solution in an overall constant. This property also holds for the $\cN=3$ SO(4) solution of \cite{Pang:2015vna}, see appendix \ref{sec:TraceMN=3}.


\subsection{$\cN=2$ M2--brane case} \label{sec:GravMassMatM2}

In order to show the relation between the massive IIA and $D=11$ cases, it is useful to first review the latter. The spectrum of KK gravitons on  the CPW AdS$_4$ solution \cite{Corrado:2001nv} of $D=11$ supergravity was computed in \cite{Klebanov:2009kp}. In that reference, this spectrum was given in terms on non-negative integers $j, p, q$ and an integer $n_r$ of either sign as
\begin{eqnarray} \label{KKGravMassM2v1}
L^2 M^2_{j,n_r,p,q} &=& 2j^2 + 2 j |n_r| + n_r^2 + 2j(p+q+3) +\tfrac13 n_r (p-q) \nonumber \\
&&  + |n_r| (3+p+q) +\tfrac19 (p^2+q^2+4pq+15p+15q) \; . 
\end{eqnarray}
In section \ref{N=2spin2sectrum} we noted the different (short) spin-2 spectra of the $\cN=2$ field theories on the M2 \cite{Benna:2008zy} and D2 \cite{Guarino:2015jca} branes. Equations (\ref{KKGravMassM2v1}) and (\ref{eq:KKGravSpectraSols}) make that difference also obvious at the level of the (both short and long) KK graviton masses. In spite of these differences, we will show that both spectra are nevertheless related.

For this purpose, it is convenient to re-express (\ref{KKGravMassM2v1}) in terms of $p$, $q$ and two new quantum numbers $n$ and $r$ defined as
\begin{eqnarray} \label{newQNms}
n = 2j + |n_r| + p +q \; , \qquad 
2r = n+ n_r -p+q  \; .
\end{eqnarray}
In the r.h.s. of the second relation, $n$ must be substituted with the expression given in the first equation. In terms of these, the spectrum of KK graviton masses (\ref{KKGravMassM2v1}) reads 
\begin{equation} \label{KKGravMassM2v2}
L^2 M^2_{n,r,p,q} = \tfrac12 n(n+6) +\tfrac12 (n-2r)^2 -\tfrac43 n(p-q) + \tfrac49 p (p+6r-3) + \tfrac49 q (q- 6r-3) -\tfrac{20}{9} pq \; .
\end{equation}
The virtue of this rewrite is that all integers $n$, $r$, $p$, $q$ now correspond to Dynkin labels, unlike $j$. In particular, $n$ is the KK level, in the sense that it labels the symmetric traceless representation of SO(8) from which the $[p,q]$ representations of SU(3) in which the spectrum is organised descend\footnote{In \cite{Klebanov:2009kp}, $j$ was referred to as the KK level. We instead dignify $n$ with that name for the reasons explained in the text.}. More precisely,
\begin{equation}
  \label{eq:SO8toSO6toU3branching}
  [n,0, 0, 0] \;
  \stackrel{\mathrm{SO}(7)_-}{\longrightarrow} \; 
 [0, 0, n] \;
  \stackrel{\mathrm{SU}(4)_-}{\longrightarrow} \; 
   \sum_{r=0}^n \, [n-r, 0 , r] \; 
  \stackrel{\mathrm{SU}(3)\times\mathrm{U}(1)_-}{\longrightarrow} \; 
   \sum_{r=0}^n  \sum_{p=0}^{n-r} \sum_{q=0}^{r} \, [p, q]_{\frac43 (p-q) + 2r - n} \; ,
\end{equation}
under the chain $\textrm{SO}(8) \supset \textrm{SO}(7)_- \supset \textrm{SU}(4)_- \supset \textrm{SU}(3) \times \textrm{U}(1)_-$. The subscripts $-$ are in line with the discussion in appendix \ref{sec:SO8triality}. The subscript on $[p,q]$ corresponds to the U$(1)_-$ R-charge $R$. In terms of the $n_r$ charge of equation (\ref{newQNms}) and the SU(3) Dynkin labels, $R$ is given by \cite{Klebanov:2009kp}
\begin{equation} \label{RchargeM2}
R = n_r + \tfrac13 (p-q)  = \tfrac43 (p-q) + 2r - n \; .
\end{equation}
At each KK level $n$, the eigenvalues (\ref{KKGravMassM2v2}) thus come in the $[p,q]$ representations of SU(3) given in (\ref{eq:SO8toSO6toU3branching}), with U$(1)_-$ R-charge (\ref{RchargeM2}). Their degeneracy is therefore
\begin{equation} \label{MassDegM2}
d_{p,q} \equiv \textrm{dim} \, [p, q]  = \tfrac12 (p+1) (q +1) (p+q+2) \; .
\end{equation}

The splitting (\ref{eq:SO8toSO6toU3branching}) allows one to read off the following ranges for the quantum numbers:
\begin{eqnarray} \label{QNrangesSO8}
n = 0, 1, 2 , \ldots \; , \qquad
r = 0 , 1 , \ldots , n \; , \qquad 
p = 0 , 1 , \ldots , n-r \; , \qquad 
q = 0 , 1 , \ldots , r \; .
\end{eqnarray}
Again, only $n$ is free to range over the non-negative integers. The ranges of the other quantum numbers are bounded. In terms of these quantum numbers, the spectrum is naturally organised KK level by KK level, as in the D2-brane cases. We can thus introduce the infinite-dimensional KK graviton mass matrix $\cM^2$. This is block-diagonal as in (\ref{Infinitemassmat}), with blocks $M^2_\n$ now of dimension $D_{n,8} \times D_{n,8}$, where $D_{n,8}$ is the dimension of the symmetric traceless representation of SO(8) given by (\ref{kSONdim}) with $k=n$ and $N=8$. At fixed KK level $n$, the quantum numbers $r$, $p$, $q$ sweep out each block $M^2_\n$. The eigenvalues of each $M^2_\n$ are the $M^2_{n,r,p,q} $ given in (\ref{KKGravMassM2v2}). For convenience, we tabulate these eigenvalues in table \ref{tab:KKGravMassSU3U1N=2SO8}, reproducing the results of \cite{Klebanov:2009kp}.

\begin{table}[h]
\centering

\resizebox{\textwidth}{!}{

\begin{tabular}{|c|l|c|c|c|c|c|c|}
\hline
$n$                 & $[p,q]_{\frac43 (p-q) + 2r - n }$          & $d_{p,q}$ & $L^2 M^2_{n,r,p,q} $        & $\Delta_{n,r,p,q}$                                          & $L^2$tr$M^2_{(n)}$          & Dual operator     & Short?   \\ \hline\hline
0                   & $[0,0]_0$                                    & 1   & 0               & 3                                                 & 0                   &$\mathcal{T}^{(0)}_{\alpha\beta}\vert_{s=2}$           & $\checkmark$     \\ \hline
\multirow{2}{*}{1}  & $[0,0]_{\pm 1}$                              & 1   & 4               & 4                                                 & \multirow{2}{*}{$\frac{56}{3}$} & $\mathcal{T}^{(0)}_{\alpha\beta}\mathcal{Z}^4\vert_{s=2}$, c.c. &$\checkmark$       \\ \cline{2-5}\cline{7-8}
                    & $[1,0]_{\frac{1}{3}},[0,1]_{-\frac{1}{3}}$   & 3   & $\frac{16}{9}$  & $\frac{1}{2} \left(\frac{\sqrt{145}}{3}+3\right)$ &                             & $\mathcal{T}^{(0)}_{\alpha\beta}\mathcal{Z}^a\vert_{s=2}$, c.c. &        \\ \hline
\multirow{6}{*}{2}  & $[0,0]_{\pm 2}$                              & 1   & 10              & 5                                                 & \multirow{6}{*}{$\frac{560}{3}$} & $\mathcal{T}^{(0)}_{\alpha\beta}(\mathcal{Z}^4)^2\vert_{s=2}$, c.c.    &$\checkmark$  \\ \cline{2-5}\cline{7-8}
                    & $[1,0]_{-\frac{2}{3}},[0,1]_{\frac{2}{3}}$   & 3   & $\frac{58}{9}$  & $\frac{1}{2} \left(\frac{\sqrt{313}}{3}+3\right)$ &                             &  $\mathcal{T}^{(0)}_{\alpha\beta}\mathcal{Z}^a\bar{\mathcal{Z}}_4\vert_{s=2}$, c.c. &        \\ \cline{2-5}\cline{7-8}
                    & $[2,0]_{\frac{2}{3}},[0,2]_{-\frac{2}{3}}$   & 6   & $\frac{34}{9}$  & $\frac{1}{2} \left(\frac{\sqrt{217}}{3}+3\right)$ &                             &  $\mathcal{T}^{(0)}_{\alpha\beta}\mathcal{Z}^{(a}\mathcal{Z}^{b)}\vert_{s=2}$, c.c. &       \\ \cline{2-5}\cline{7-8}
                    & $[1,0]_{\frac{4}{3}},[0,1]_{-\frac{4}{3}}$   & 3   & $\frac{64}{9}$  & $\frac{1}{2} \left(\frac{\sqrt{337}}{3}+3\right)$ &                             & $\mathcal{T}^{(0)}_{\alpha\beta}\mathcal{Z}^a {\mathcal{Z}}^4\vert_{s=2}$ , c.c. &        \\ \cline{2-5}\cline{7-8}
                    & $[0,0]_0$                                    & 1   & 8               & $\frac{1}{2} \left(\sqrt{41}+3\right)$            &                               & $\mathcal{T}^{(0)}_{\alpha\beta}(1-4a^2\mathcal{Z}^4\bar{\mathcal{Z}}_4)\vert_{s=2}$  &      \\ \cline{2-5}\cline{7-8}
                    & $[1,1]_0$                                    & 8   & 4               & 4                                                 &                                &  $\mathcal{T}^{(0)}_{\alpha\beta}(\mathcal{Z}^a \bar{{\mathcal{Z}}_b}-\frac{1}{3}\delta^{a}_{b}\mathcal{Z}^c\bar{\mathcal{Z}}_c)\vert_{s=2}$&     \\ \hline
\multirow{10}{*}{3} & $[0,0]_{\pm 3}$                              & 1   & 18              & 6                                                 & \multirow{10}{*}{1008}          & $\mathcal{T}^{(0)}_{\alpha\beta}(\mathcal{Z}^4)^3\vert_{s=2}$, c.c. & $\checkmark$  \\ \cline{2-5}\cline{7-8}
                    & $[1,0]_{-\frac{5}{3}},[0,1]_{\frac{5}{3}}$   & 3   & $\frac{118}{9}$ & $\frac{1}{2} \left(\frac{\sqrt{553}}{3}+3\right)$ &                    & $\mathcal{T}^{(0)}_{\alpha\beta}\mathcal{Z}^a(\bar{\mathcal{Z}}_4)^2\vert_{s=2}$, c.c &                 \\ \cline{2-5}\cline{7-8}
                    & $[2,0]_{-\frac{1}{3}},[0,2]_{\frac{1}{3}}$   & 6   & $\frac{82}{9}$  & $\frac{1}{2} \left(\frac{\sqrt{409}}{3}+3\right)$ &                      & $\mathcal{T}^{(0)}_{\alpha\beta}\mathcal{Z}^{(a}\mathcal{Z}^{b)}(\bar{\mathcal{Z}}_4)\vert_{s=2}$, c.c. &                 \\ \cline{2-5}\cline{7-8}
                    & $[3,0]_{1},[0,3]_{-1}$                       & 10  & 6               & $\frac{1}{2} \left(\sqrt{33}+3\right)$            &                              & $\mathcal{T}^{(0)}_{\alpha\beta}\mathcal{Z}^{(a}\mathcal{Z}^b\mathcal{Z}^{c)}\vert_{s=2}$, c.c. &       \\ \cline{2-5}\cline{7-8}
                    & $[0,0]_{\pm 1}$                              & 1   & 14              & $\frac{1}{2} \left(\sqrt{65}+3\right)$            &                                 & $\mathcal{T}^{(0)}_{\alpha\beta}(2-5a^2\mathcal{Z}^4\bar{\mathcal{Z}}_4)\mathcal{Z}^4\vert_{s=2}$, c.c. &      \\ \cline{2-5}\cline{7-8}
                    & $[1,0]_{\frac{7}{3}},[0,1]_{-\frac{7}{3}}$   & 3   & $\frac{130}{9}$ & $\frac{1}{2} \left(\frac{\sqrt{601}}{3}+3\right)$ &                & $\mathcal{T}^{(0)}_{\alpha\beta}\mathcal{Z}^a(\mathcal{Z}^4)^2\vert_{s=2}$, c.c. &                       \\ \cline{2-5}\cline{7-8}
                    & $[1,0]_{\frac{1}{3}},[0,1]_{-\frac{1}{3}}$   & 3   & $\frac{106}{9}$ & $\frac{1}{2} \left(\frac{\sqrt{505}}{3}+3\right)$ &                  & $\mathcal{T}^{(0)}_{\alpha\beta}\mathcal{Z}^a(1-5a^2\mathcal{Z}^4\bar{\mathcal{Z}}_4)\vert_{s=2}$, c.c. &                     \\ \cline{2-5}\cline{7-8}
                    & $[1,1]_{\pm 1}$                              & 8   & 10              & 5                                                 &                              &$\mathcal{T}^{(0)}_{\alpha\beta}(\mathcal{Z}^a \bar{{\mathcal{Z}}_b}-\frac{1}{3}\delta^{a}_{b}\mathcal{Z}^c\bar{\mathcal{Z}}_c)\mathcal{Z}^4\vert_{s=2}$, c.c. &         \\ \cline{2-5}\cline{7-8}
                    & $[2,0]_{\frac{5}{3}}, [0,2]_{-\frac{5}{3}} $ & 6   & $\frac{94}{9}$  & $\frac{1}{2} \left(\frac{\sqrt{457}}{3}+3\right)$ &                 &  $\mathcal{T}^{(0)}_{\alpha\beta}\mathcal{Z}^{(a}\mathcal{Z}^{b)}{\mathcal{Z}}^4\vert_{s=2}$, c.c&                      \\ \cline{2-5}\cline{7-8}
                    & $[2,1]_{\frac{1}{3}}, [1,2]_{-\frac{1}{3}} $ & 15  & $\frac{58}{9}$  & $\frac{1}{2} \left(\frac{\sqrt{313}}{3}+3\right)$ &                 & $\mathcal{T}^{(0)}_{\alpha\beta}\mathcal{Z}^{(a}\mathcal{Z}^{b)}\bar{\mathcal{Z}}_c-\delta^{(a}_{c}\mathcal{Z}^{b)}\mathcal{Z}^d\bar{\mathcal{Z}}_d)\vert_{s=2}$, c.c. &                      \\ \hline
\end{tabular}

}
\caption{\footnotesize{The spectrum of KK gravitons on the CPW $\cN=2$ $\textrm{SU}(3) \times \textrm{U}(1)$--invariant solution \cite{Corrado:2001nv} of $D=11$ supergravity up to KK level $n=3$, reproduced from \cite{Klebanov:2009kp}. For each state, its $\textrm{SU}(3) \times \textrm{U}(1)$ charges (\ref{eq:SO8toSO6toU3branching}), degeneracy (\ref{MassDegM2}), mass (\ref{KKGravMassM2v2}) and dimension computed from (\ref{DeltaM}) is given.  The trace of the mass matrix (\ref{trMngeneralM2}) at level $n$ is also given, and the schematic form of the dual  single-trace spin-2 operators. Checked (unchecked) states belong to short (long) graviton supermultiplets of OSp$(4|2)$.}\normalsize}
\label{tab:KKGravMassSU3U1N=2SO8}
\end{table}

We can now compute the trace of each block  $M^2_\n$ as we did in section \ref{sec:GravMassMatD2} for the D2-brane case. Using the eigenvalues $M^2_{n,r,p,q} $ given in (\ref{KKGravMassM2v2}),  the degeneracies $d_{p,q}$ given in (\ref{MassDegM2}), and summing over the ranges (\ref{QNrangesSO8}) at fixed KK level $n$, we obtain
\begin{eqnarray} \label{trMngeneralM2}
L^2 \; \tr \, M^2_\n =  L^2 \;  \sum_{r=0}^n  \sum_{p=0}^{n-r} \sum_{q=0}^{r} \, M^2_{n, r , p, q } \, d_{p , q} = \tfrac{56}{3}  \, D_{n-1 , \, 10} \; .
\end{eqnarray}
The result is an 8th order polynomial in $n$ which, like for the D2-brane cases discussed in section \ref{sec:GravMassMatD2}, can be compactly written using the formula (\ref{kSONdim}) for the symmetric traceless representation at Dynkin label $k=n-1$ of SO$(N)$, now with $N=10$. Again, we do not have an explanation as to why the result can be expressed in terms of the dimension of this representation of SO$(10)$ or any other group, and simply employ the notation $D_{n-1 , \, 10}$ to express the result in a compact way. However, this is not a coincidence: this property is shared by the uplifts of other critical points of SO(8)--gauged supergravity. This will be shown elsewhere, but it is readily seen for the $\cN=8$ SO(8)--invariant critical point, which uplifts to the Freund-Rubin vacuum \cite{Freund:1980xh} of $D=11$ supergravity. In this case, the mass of the KK graviton at level $n$ is $L^2 M_n^2 = \tfrac14 n(n+6)$ (see \cite{Duff:1986hr}) and occurs with degeneracy $D_{n,8} = \textrm{dim} [n,0,0,0]$, so that each block $M_\n$ in the mass matrix $\cM^2$ equals that eigenvalue times the identity matrix of dimension $D_{n,8}$. Thus, for the $\cN=8$ SO(8) critical point,
\begin{equation}
L^2 \; \tr \, M^2_\n = \tfrac14 n(n+6) \cdot \tfrac{1}{6!} (2n+6) (n+5) (n+4) (n+3) (n+2) (n+1) = 14  \, D_{n-1 , \, 10} \; .
\end{equation}

\subsection{Universality of the D2 and M2 graviton mass matrix traces}

Observe that the coefficient $\frac{56}{3}$ for the $D=11$ result (\ref{trMngeneralM2}) matches the coefficient for the massive IIA case given in the first line of (\ref{eq:TrSols}). This translates into a relation between both mass matrix traces. To see this, we need the following property
\begin{equation} \label{eq:DimRel}
D_{n,N-1} = D_{n,N} -  D_{n-1, \, N}   \; , 
\end{equation}
of the dimension of the symmetric traceless representation of the orthogonal group, which easily follows from (\ref{kSONdim}). The property (\ref{eq:DimRel}) and the fact that both mass matrix traces have the same coefficient implies the universality relation
\begin{eqnarray} \label{tracerelation}
L^2_\textrm{D2} \;  \tr \, M^2_{\n \, \textrm{D2}} \,= \, L^2_\textrm{M2} \; \Big( \tr \, M^2_{\n \, \textrm{M2}} - \tr \, M^2_{(n-1) \, \textrm{M2}}  \Big) \; , \qquad n= 1 , 2 , \ldots  
\end{eqnarray}
Here, we have added labels D2 and M2 to the quantities in (\ref{eq:TrSols}) and (\ref{trMngeneralM2}) corresponding to the solutions of massive IIA \cite{Guarino:2015jca} and $D=11$ \cite{Corrado:2001nv}, respectively. The relation (\ref{eq:DimRel}) implies that the traces on both sides of equation (\ref{tracerelation}) are effectively taken over the same number of states. The traces tabulated in tables \ref{tab:KKGravMassSU3U1N=2ISO7} and \ref{tab:KKGravMassSU3U1N=2SO8} are useful for a quick check of the relation (\ref{tracerelation}) up to KK level $n=3$.

Here we have only shown the universality relation (\ref{tracerelation}) to hold for the $\cN=2$ $\textrm{SU}(3)\times \textrm{U}(1)$ solutions of massive IIA \cite{Guarino:2015jca}  and $D=11$ \cite{Corrado:2001nv} supergravity, which respectively uplift from the vacua of $D=4$ $\cN=8$ dyonic ISO(7) \cite{Guarino:2015jca} and electric SO(8) \cite{Warner:1983vz} supergravities with that symmetry. However, we have checked that this is not an isolated case: the relation (\ref{tracerelation}) also holds for the massive IIA and $D=11$ uplifts of other pairs of vacua of these $\cN=8$ supergravities with the same symmetries. Further details will be given elsewhere. Thus, while the strong, eigenvalue by eigenvalue universality of the $n=0$ KK mass spectra is broken at higher KK levels, a certain form of universality is nevertheless still preserved at the level of the KK graviton mass matrix traces.

\section{$\cN=2$ spectrum at $n=0$ KK level} \label{subsec:conjFullKKn=0}

We have seen how the universality of the $n=0$ KK mass spectrum of the AdS$_4$ solutions of massive IIA and $D=11$ supergravity under consideration is lost at higher KK levels (though still maintained in a weaker form). The higher-dimensional origin, the compactification manifolds and the dual field theories are different, so differences in the spectrum were expected to arise. Here, we would like to enquire if these differences actually manifest themselves even at KK level $n=0$, in spite of the universality of supergravity masses at this level. It turns out that they do, in a very subtle way. We will again focus on the $\cN=2$ $\textrm{SU}(3) \times \textrm{U}(1)$--invariant vacua of the SO(8) and dyonic ISO(7) supergravities and show that the spectra within their respective $\cN=8$ supergravities share the same supermultiplet structure, with a few different R-symmetry and conformal dimension assignments.

\begin{table}[]
\begin{center}
\begin{tabular}{|c|c|c|c|c|c|c|c|c|c|} \hline
Spin & $\grSO(7)$ & \multicolumn{8}{c|}{$\grSU(3)_{\grU(1)}$} \\ \hline
%
%
$2$ & $\rep{1}$ & $\rep{1}_{0}$ &&&&&&&   \\ \hline
%
%
$\frac{3}{2}$ & $\rep{8}$ & $\rep{1}_{+1}$ & & $\rep{3}_{\et}$ & $\rep{\bar{3}}_{-\et}$ &  &  && \\
&&$\rep{1}_{-1}$ & &&&&&&  \\  \hline
%
%
$1$ & $\rep{21}+\rep{7}$ &  $\rep{1}_{0}$ & $\rep{8}_{0}$& $\rep{3}_{\eft}$ & $\rep{\bar{3}}_{-\eft}$ && & $\rep{1}_{0}$ & \\
&&&& $\rep{3}_{-\ett}$ & $\rep{\bar{3}}_{\ett}$ &&&& \\
&&&& $\rep{3}_{-\ett}$ & $\rep{\bar{3}}_{\ett}$  && && \\ \hline
%
%
$\frac{1}{2}$ & $\rep{48}+\rep{8}$ & &$\rep{8}_{+1}$  & $\rep{3}_{\et}$ & $\rep{\bar{3}}_{-\et}$ & $\rep{6}_{-\et}$ & $\rep{\bar{6}}_{\et}$ & $\rep{1}_{-1}$ & $\rep{3}_{\et}$ \\  
&& &$\rep{8}_{-1}$ &$\rep{3}_{\et}$ & $\rep{\bar{3}}_{-\et}$ &&  & $\rep{1}_{+1}$ & $\rep{\bar{3}}_{-\et}$ \\ 
&&&   & $\rep{3}_{-\evt}$ & $\rep{\bar{3}}_{\evt}$ & & & $\rep{1}_{-1}$ & \\ 
&&&&&&& & $\rep{1}_{+1}$ & \\ \hline 
%
%
$0^+$ & $\rep{27}+\rep{7}+\rep{1}$ & & $\rep{8}_{0}$ &  $\rep{3}_{-\ett}$ & $\rep{\bar{3}}_{\ett}$  & $\rep{6}_{-\eft}$ & $\rep{\bar{6}}_{\eft}$ &  $\rep{1}_{0}$ & $\rep{3}_{-\ett}$ \\
&&&&&&&& $\rep{1}_{0}$ & $\rep{\bar{3}}_{\ett}$ \\
&&&&&&&&& $\rep{1}_0$ \\ \hline
%
%
$0^-$ & $\rep{35}$ & &$\rep{8}_{0}$ & && $\rep{6}_{\ett}$ & $\rep{\bar{6}}_{-\ett}$ & $\rep{1}_{+2}$ &  $\rep{3}_{\eft}$ \\
&&&&&&&& $\rep{1}_{0}$ & $\rep{3}_{-\ett}$ \\
&&&&&&&& $\rep{1}_{-2}$ & $\rep{\bar{3}}_{-\eft}$ \\
&&&&&&&&& $\rep{\bar{3}}_{\ett}$ \\ \hline
\multicolumn{2}{l|}{} &
\rotatebox{90}{\mbox{Massless graviton\;}} &
\rotatebox{90}{\mbox{Massless vector\;}} &
\rotatebox{90}{\mbox{Massive short gravitino\;}} &
\rotatebox{90}{\mbox{Massive short gravitino\;}} &
\rotatebox{90}{\mbox{Massive hyper\;}} &
\rotatebox{90}{\mbox{Massive hyper\;}} &
\rotatebox{90}{\mbox{Massive vector\;}} &
\rotatebox{90}{\mbox{eaten\;}} \\  \cline{3-10}
\end{tabular}
\caption{The OSp$(4|2) \times \textrm{SU}(3) $ spectrum of the $\cN=2$ solution of $\cN=8$ dyonic ISO(7) supergravity within the $\cN=8$ theory. This coincides with scenario II of KKM \cite{Klebanov:2008vq}.}
\label{tab:n=0SU3U1D2}
\end{center}
\end{table}

The mass spectrum of the  $\cN=2$ $\textrm{SU}(3) \times \textrm{U}(1)$ vacuum \cite{Warner:1983vz} of electrically-gauged SO(8) supergravity \cite{deWit:1982ig} was computed and allocated into OSp$(4|2) \times \textrm{SU}(3)$ representations by Nicolai and Warner \cite{Nicolai:1985hs}. They noted that the U(1) R-symmetry could be embedded in SO(8) in two possible ways, which they argued to be essentially equivalent for their purposes. 
More recently, this question was re-examined by Klebanov, Klose and Murugan (KKM) \cite{Klebanov:2008vq}. KKM renamed these two possible U(1) embeddings as scenarios I and II and realised that, while both scenarios led to the same masses for all fields within $\cN=8$ SO(8) supergravity, they led to small differences in the R-charge and conformal dimensions of a few fields. Except for these minor differences, the supermultiplet structure implied by both scenarios was  found to be identical. A KK graviton analysis \cite{Klebanov:2009kp} confirmed the choice of \cite{Nicolai:1985hs}, KKM's scenario I  \cite{Klebanov:2008vq}, as the correct one describing the spectrum of the $\cN=2$ vacuum \cite{Warner:1983vz} of $\cN=8$ SO(8) supergravity within this theory. We will now show that the spectrum of the $\cN=2$ vacuum \cite{Guarino:2016ynd,Guarino:2015qaa} of $\cN=8$ dyonic ISO(7) supergravity within the latter theory turns out to realise KKM's scenario II. In appendix \ref{sec:SO8triality} we also show that the existence of these two scenarios and their associated R-charge assignments have an origin in SO(8) triality.

For the $\cN=2$ vacuum of the dyonic ISO(7) supergravity, the embedding of the residual $\textrm{SU}(3) \times \textrm{U}(1) $ symmetry  into the compact $ \textrm{SO}(7)$ subgroup of the ISO(7) gauge group is unique. The fundamental, $\textbf{7} \equiv [1,0,0]$, and spinor, $\textbf{8} \equiv [0,0,1]$, representations of SO(7) branch under $\textrm{SO}(7) \supset \textrm{SO}(6) \supset \textrm{SU}(3) \times \textrm{U}(1)$ uniquely as 
\begin{eqnarray}
  \label{eq:7ofSO7toSO6toU3branching}
  && \textbf{7} \;
  \stackrel{\mathrm{SO}(6)}{\longrightarrow} \; 
   \textbf{6} + \textbf{1} 
  \stackrel{\mathrm{SU}(3)\times\mathrm{U}(1)}{\longrightarrow} \; 
   \big( \textbf{3}_{-\frac23} + \overline{\textbf{3}}_{+\frac23} \big) +  \textbf{1}_0 \; , \\
  \label{eq:8ofSO7toSO6toU3branching}
  && \textbf{8} \;
  \stackrel{\mathrm{SO}(6)}{\longrightarrow} \; 
   \textbf{4} + \overline{\textbf{4}} 
  \stackrel{\mathrm{SU}(3)\times\mathrm{U}(1)}{\longrightarrow} \; 
   \big( \textbf{3}_{\frac13} + \textbf{1}_{-1} \big) + \big( \overline{\textbf{3}}_{-\frac13} +  \textbf{1}_{+1} \big) \; .
\end{eqnarray}
The branching (\ref{eq:7ofSO7toSO6toU3branching}) is of course (\ref{eq:SO7toSO6toU3branching}) with $n=1$. It is natural to assign the (electric) vectors, scalars and pseudoscalars of the $\cN=8$ ISO(7) theory to the $\textbf{21} + \textbf{7}$, the $\textbf{27} + \textbf{7}+ \textbf{1}$ and the $\textbf{35}$ representations of SO(7), respectively, and the gravitino and spin $1/2$ fermions to the $\textbf{8}$ and $\textbf{48}+\rep{8}$, see appendix \ref{sec:SO8triality} for a justification. Tensoring (\ref{eq:7ofSO7toSO6toU3branching}) and (\ref{eq:8ofSO7toSO6toU3branching}) with themselves and (anti)symmetrising appropriately, we determine how these SO(7) representations decompose under $\textrm{SU}(3) \times \textrm{U}(1)$. Finally, we group up fields in the same SU(3) representations into OSp$(4|2)$ multiplets \cite{Freedman:1983na,Ceresole:1984hr,Nicolai:1985hs} (see also appendix A of \cite{Klebanov:2008vq}). In this way, we obtain the OSp$(4|2) \times \textrm{SU}(3)$ breakdown of the $\cN=8$ supergravity fields at the $\cN=2$ point: see table \ref{tab:n=0SU3U1D2}. This table exactly matches table 5 of \cite{Klebanov:2008vq}, corresponding to KKM's scenario II, with the choice $\varepsilon = +1$ for the arbitrary R-charge sign $\varepsilon$. We kindly borrow their format for ease of comparison. The only technical difference with KKM's scenario II is that, in their case, the $\cN=8$ fields naturally group up in SO(8) representations before branching into $\textrm{SU}(3) \times \textrm{U}(1)$, whereas in the present case the $\cN=8$ fields fill out instead the SO(7) representations described above and recorded in table \ref{tab:n=0SU3U1D2}. 

The structure of $\textrm{OSp}(2|4) \times \textrm{SU}(3)$ representations is identical for both scenarios I and II, except that the hypermultiplet in the $\bm{6}$ of SU(3) has U(1) R-charge $R = \frac23$ in the first case and $R = -\frac43$ in the second. The conjugate hypermultiplets, in the $\bm{\bar{6}}$ of SU(3), have the same R-charges with opposite sign. This is the only difference as far as the allocation of $\cN=8$ supergravity fields into supermultiplets is concerned. There are a few other differences, though. The long vector multiplet of scenario I contains three scalars with R-charges $0$, $\pm2$, and two neutral pseudoscalars. In scenario II, the role of scalars and pseudoscalars within the long vector multiplet is exchanged. In \cite{Guarino:2015qaa} (and in section \ref{sec:KKGravitonsSU3} above), these SU(3)-singlet scalars were denoted by $\phi$, $\varphi$, the neutral pseudoscalar by $\chi$ and the charged pseudoscalars by $\zeta$, $\tilde{\zeta}$. There are also differences in the structure of Goldstone bosons. For example, in scenario I (II), the massive vector eats an R-neutral SU(3) singlet pseudoscalar (scalar). This eaten scalar was denoted by $a$ in \cite{Guarino:2015qaa}. Incidentally, the $D=4$ supergravity describing the full non-linear interactions of the (linearised) SU(3)-singlet fields in table \ref{tab:n=0SU3U1D2} was constructed in \cite{Guarino:2015qaa} and uplifted to massive type IIA in \cite{Varela:2015uca}. In particular, equation (\ref{VSU3}) above is the full, non-linear potential for the SU(3)-singlet scalars and pseudoscalars, and (\ref{KKSU3sectorinIIA}) describes their embedding into the ten-dimensional metric. The analogue $D=4$ supergravity containing the full non-linear interactions of the SU(3) singlets of scenario I was constructed in \cite{Bobev:2010ib}.

\begin{table}[t]
\centering
\resizebox{\textwidth}{!}{%
\begin{tabular}{|l|c|c|c|l|}
\hline
scalar/pseudoscalar                                                                                                                                                     & $\textrm{SU}(3)_{\textrm{U}(1)}$   & $M^2L^2$        & $\Delta$                & Osp($4|2$) multiplet \\ \hline\hline
$Z^a\bar{Z}_b-\frac{1}{3}\delta^a_bZ^c\bar{Z}_c$                                                                                                            & $\mathbf{8}_0$            & $-2$              & 1                       & massless vector      \\ \hline
$Z^a\bar{Z}_4$                                                                                                                                              & $\mathbf{3}_{-2/3}$       & $-\frac{14}{9}$ & $\frac{7}{3}$           & short gravitino      \\ \hline
$\bar{Z}_a Z^4$                                                                                                                                             & $\mathbf{\bar{3}}_{2/3}$  & $-\frac{14}{9}$ & $\frac{7}{3}$           & short gravitino      \\ \hline
$Z^{(a}Z^{b)}$                                                                                                                                              & $\mathbf{6}_{-4/3}$       & $-\frac{20}{9}$ & $\frac{4}{3}$           & hypermultiplet      \\ \hline
$\bar{Z}_{(a}\bar{Z}_{b)}$                                                                                                                                  & $\mathbf{\bar{6}}_{4/3}$  & $-\frac{20}{9}$ & $\frac{4}{3}$           & hypermultiplet      \\ \hline
$Z^{a}\bar{Z}_a-3Z^4 \bar{Z}_4$                                                                                                                                    & $\mathbf{1}_0$            & $3-\sqrt{17}$   & $\frac{1+\sqrt{17}}{2}$ & long vector          \\ \hline
Re($Z^4Z^4$)                                                                                                                                                & $\mathbf{1}_0$            & $3+\sqrt{17}$   & $\frac{5+\sqrt{17}}{2}$ & long vector          \\ \hline
Im($Z^4Z^4$)                                                                                                                                                & $\mathbf{1}_0$            & 0               & -                       & eaten                \\ \hline
$Z^aZ^4$                                                                                                                                                    & $\mathbf{3}_{-2/3}$       & 0               & -                       & eaten                \\ \hline
$\bar{Z}_a\bar{Z}_4$                                                                                                                                        & $\mathbf{\bar{3}}_{2/3}$  & 0               & -                       & eaten                \\ \hline\hline
$dZ^a\wedge d\bar{Z}_b\wedge dZ^4\wedge d\bar{Z}_4 -\frac{1}{4}\epsilon^{acd}\epsilon_{bef}d\bar{Z}_c\wedge d\bar{Z}_d\wedge dZ^e \wedge dZ^f-$ trace             & $\mathbf{8}_0$            & $-2$              & 2                       & massless vector      \\ \hline
$\epsilon^{cd(a} dZ^{b)} \wedge d\bar{Z}_c\wedge d\bar{Z}_d\wedge d{Z}^4$                                                                                     & $\mathbf{6}_{2/3}$        & $-\frac{14}{9}$  & $\frac{7}{3}$           & hypermultiplet      \\ \hline
$\epsilon_{cd(a} d\bar{Z}_{b)} \wedge d{Z}^c \wedge d{Z}^d\wedge d\bar{Z}_4$                                                                                & $\mathbf{\bar{6}}_{-2/3}$ & $-\frac{14}{9}$  & $\frac{7}{3}$           & hypermultiplet      \\ \hline
$\epsilon_{abc}dZ^a\wedge dZ^b\wedge dZ^c\wedge d{Z}^4$                                                                                                         & $\mathbf{1}_{-2}$         & $2  $            & $\frac{3+\sqrt{17}}{2}$ & long vector          \\ \hline
$\epsilon^{abc}d\bar{Z}_a\wedge d\bar{Z}_b\wedge d\bar{Z}_c \wedge d\bar{Z}_4$                                                                          & $\mathbf{1}_{2}$         &  $2$           & $\frac{3+\sqrt{17}}{2}$ & long vector          \\ \hline
$dZ^a\wedge d\bar{Z}_a\wedge dZ^b\wedge d\bar{Z}_b+2 dZ^a\wedge d\bar{Z}_a\wedge dZ^4\wedge d\bar{Z}_4$                                                                                             & $\mathbf{1}_0$          & $ 2  $            & $\frac{3+\sqrt{17}}{2}$ & long vector          \\ \hline
$dZ^a \wedge dZ^b\wedge d\bar{Z}_b\wedge d{Z}^4$                                                                                                              & $\mathbf{3}_{-2/3}$       & 0               & -                       & eaten                \\ \hline
$d\bar{Z}_a \wedge d\bar{Z}_b \wedge dZ^b \wedge d\bar{Z}_4$                                                                                                      & $\mathbf{\bar{3}}_{2/3}$  & 0               & -                       & eaten                \\ \hline
$\epsilon^{bcd}dZ^a\wedge d\bar{Z}_b\wedge d\bar{Z}_c\wedge d\bar{Z}_d+3\epsilon^{abc}d\bar{Z}_b \wedge d\bar{Z}_c \wedge dZ^4 \wedge d\bar{Z}_4$ & $\mathbf{3}_{4/3}$        & 0               & -                       & eaten                \\ \hline
$\epsilon_{bcd}d\bar{Z}_a\wedge d{Z}^b\wedge d{Z}^c\wedge d{Z}^d+3\epsilon_{abc}d{Z}^b \wedge d{Z}^c \wedge d\bar{Z}_4 \wedge d{Z}^4$             & $\mathbf{\bar{3}}_{-4/3}$ & 0               & -                       & eaten                \\ \hline
\end{tabular}%
}
\caption{
\footnotesize{The spectrum of scalars and pseudoscalars  on the $\cN=2$ $\textrm{SU}(3) \times \textrm{U}(1)$--invariant solution of massive type IIA \cite{Guarino:2015jca} at KK level $n=0$. The OSp$(4|2)$ supermultiplet to which each these belong according to table \ref{tab:n=0SU3U1D2} is indicated.}}
\label{tab:n=0SU3U1D2masses}
\end{table}

 The masses of the $\cN=8$ ISO(7) supergravity fields at the $\cN=2$ point were computed in \cite{Guarino:2015qaa}. Now, we can partially reproduce this mass spectrum from group theory. All fields except those falling in the SU(3)--singlet multiplet that does not contain the graviton fill in short representations of OSp$(4|2)$. Thus, their U(1) R-charges fix their conformal dimensions (and the latter then determine their masses, as always). The SU(3)--singlets other than those that fill out the massless graviton multiplet belong to a long vector multiplet. For this reason, their dimensions are not fixed by the R-symmetry. For these fields, we merely import their masses from \cite{Guarino:2015qaa}. We reproduce the scalar and pseudoscalar masses around the $\cN=2$ vacuum in table \ref{tab:n=0SU3U1D2masses}. The table includes for convenience the schematic form of the mass eigenstates in the SL(8) basis of the $\cN=8$ supergravity (note, however, that additional mixings might occur). In this basis, the relation between $\cN=8$ scalars, $\mathbb{R}^7$ coordinates transverse to the D2-branes, and the dual operators is most transparent. The pseudoscalar mass eigenstates are included for completeness, even though their expressions in the SL(8) basis in terms of selfdual four-forms is not too enlightening. A triality rotation relates these to fermion bilinears of the boundary theory.

We conclude this section with a discussion of the form of the field theory operators in protected supermultiplets that are dual to $n=0$ KK modes. Some of these can be inferred from tables \ref{tab:n=0SU3U1D2} and \ref{tab:n=0SU3U1D2masses}. As we already discussed in section \ref{N=2spin2sectrum}, the massless graviton multiplet is dual to the stress-energy tensor superfield (\ref{SETsuperfield}). The octet massless vector multiplet is dual to the conserved global SU(3) supercurrent multiplet ${\cal J}_a ^{\0 b}$, whose scalar component can be read off from table \ref{tab:n=0SU3U1D2masses} to be given by the operator
\begin{equation}
\tr \, Z^a\bar{Z}_b-\tfrac{1}{3}\delta^a_b \, \tr \,  Z^c\bar{Z}_c \; .
\end{equation}
The dimension of this operator is fixed to $\Delta = 1$, in agreement with the supergravity result of table \ref{tab:n=0SU3U1D2masses}, because the spin-one component of ${\cal J}_a ^{\0 b}$, the conserved SU(3) global current 
$
  J_{\mu a}^{(0)b} =  \tr \,  \bar{Z}_a \stackrel{\leftrightarrow}{\partial}_\mu Z^b - \frac{1}{3}\delta_a^b  \tr \,  \bar{Z}_c \stackrel{\leftrightarrow}{\partial}_\mu  Z^c \; 
$,
must have protected classical dimension $\Delta = 2$. From table \ref{tab:n=0SU3U1D2masses}, the sextet hypermultiplets can be seen to be dual to mass terms for the chiral and antichiral superfields $\cZ^a$, $\bar{\cZ}_{a}$, 
\begin{equation} \label{masstermsChirals}
\tr \, \cZ^{(a} \cZ^{b)} \quad , \qquad 
\tr \, \bar{\cZ}_{(a} \bar{\cZ}_{b)}  \; .
\end{equation}
According to table \ref{tab:n=0SU3U1D2masses}, these have exactly the protected dimension, $\Delta = \frac43$, that follows from the assignment (\ref{DeltaandRofZ}) for the chirals. In the analogue M2-brane $\cN=2$ field theory \cite{Benna:2008zy}, these mass terms have instead protected dimension $\Delta = \frac23$, in agreement with the relevant dimension assignment (\ref{DeltaandRofZM2}). Finally, for the fields dual to the the short gravitini we propose  the following fermionic superfields
\begin{equation}
\tr \, \epsilon_{abc}\cZ^{b} D_{\alpha}\cZ^c \; , \qquad 
\tr \, \epsilon^{abc} \bar{\cZ}_{b} \bar{D}_{\alpha} \bar{\cZ}_c \; , 
\end{equation}
These respectively transform in the $\mathbf{\bar{3}}_{-\frac{1}{3}}$ and $\mathbf{3}_{\frac{1}{3}}$ and have scaling dimension $\Delta=\frac{11}{6}$. These assignments correctly reproduce the assignments $R=\frac{2}{3}$ and $\Delta=\frac{7}{3}$ recorded in table \ref{tab:n=0SU3U1D2masses} for the scalar components in these superfields.


\section{Final comments}

Pairs of AdS$_4 \times S^6$ and AdS$_4 \times S^7$ solutions of massive IIA and $D=11$ supergravity that respectively uplift from pairs of vacua of $\cN=8$ dyonic ISO(7) and electric SO(8) supergravity with the same symmetries, exhibit an identical mass spectrum at KK level $n=0$. The $n=0$ KK level mass spectrum is thus universal, and insensitive to the $S^6$ or $S^7$ compactification manifold and to the massive IIA or $D=11$ origin. We have shown that this universality is lost at higher KK levels, $n \geq 1$, as expected: the masses of the higher KK modes do differ. We have seen this explicitly for the spectrum of KK gravitons above the $\cN=2$ AdS$_4$ solutions of massive IIA  \cite{Guarino:2015jca} and CPW \cite{Corrado:2001nv} of $D=11$ supergravity. The spectra nevertheless still exhibit a weaker form of universality: the traces of the KK graviton mass matrices match KK level by KK level. At least for the supersymmetric solutions, a similar KK level by KK level match might be enforced upon the traces of mass matrices of the fields of spin $s<2$ by supersymmetry. Checking this explicitly seems a complicated task because, in principle, the entire KK spectrum about those solutions should be calculated first, and that is a difficult problem.

Since the masses of higher KK modes differ, the spectrum of dual single trace operators obviously differs too. We have illustrated this explicitly for the spectrum of protected spin-2 operators of the $\cN=2$ infrared Chern-Simons field theories on the D2 \cite{Guarino:2015jca} and M2 \cite{Benna:2008zy} branes dual to the AdS$_4$ solutions of massive IIA \cite{Guarino:2015jca} and $D=11$ \cite{Corrado:2001nv} supergravity. In fact, for these field theories, one does not need to go beyond the $n=0$ KK level to start noticing differences in the spectrum of dual single trace operators, even if the supergravity masses are identical at this level. Already at KK level $n=0$, subtle differences arise. These are due to different R-symmetry (and conformal symmetry) assignments in either case, coming in turn from different U(1) R-symmetry embeddings in the respective $\cN=8$ supergravities. These two possible U(1) embeddings were called scenarios I and II by KKM \cite{Klebanov:2008vq}, and can be understood in terms of SO(8) triality. The spectrum of the CPW solution \cite{Corrado:2001nv} realises scenario I at lowest \cite{Nicolai:1985hs} and higher \cite{Klebanov:2008vq,Klebanov:2009kp} KK levels. The massive IIA $\cN=2$ AdS$_4$ solution of \cite{Guarino:2015jca} turns out to realise scenario II at lowest, $n=0$, KK level.

Of course, scenario II should also be realised up the KK tower for the massive IIA $\cN=2$ solution of \cite{Guarino:2015jca}. Indeed, the group theory method of KKM \cite{Klebanov:2008vq} does find a series of short graviton multiplets with precisely the $\textrm{SU}(3) \times \textrm{U}(1)$ charges of (\ref{SU3U1chargesConDim}): see tables 22 and 23 of \cite{Klebanov:2008vq}. A puzzle however arises, because group theory also predicts series of multiplets with no KK interpretation. For example, KKM note an infinite series of $\textrm{SU}(3) \times \textrm{U}(1)$--neutral massless gravitons. However, these offending representations can be argued to be projected out from the physical spectrum. We will return to these questions in the future.

\section*{Acknowledgements}
We would like to thank Kanghoon Lee, Achilleas Passias and Alejandro Rosabal for discussions. YP is supported by an Alexander von Humboldt fellowship. OV is supported by NSF grant PHY-1720364 and, partially, by grant FPA2015-65480-P (MINECO/FEDER UE) from the Spanish Government.  
 
\appendix

\section{Mass operator} \label{app:massop}

Here we give some details on the derivation of the eigenvalue equation (\ref{SU3eqMassGrav}) from the general equation (\ref{FPeq}) evaluated on the geometry (\ref{KKSU3sectorinIIA}). We start by noting that the latter line element can be formally written as 
{\setlength\arraycolsep{2pt}
\begin{eqnarray} \label{KKSU3sectorinIIAbis}
d\bar{s}_{6}^2 =  L^{-2} g^{-2} \,   \big[ \;  e^{-2\phi+\varphi}   X^{-1}  d\alpha^2      +   d\tilde{s}_5^2 \, \big] \; , 
\end{eqnarray}
}where, for each constant value of $\alpha$ within its range (\ref{anglerange}), the five-dimensional geometry
\begin{eqnarray} \label{5Dtilde}
d\tilde{s}_5^2  =  d\tilde{s}^2 ( \mathbb{CP}^2 ) + (d\tilde{\psi} + \tilde{\sigma})^2 \; , 
\end{eqnarray}
with
\begin{eqnarray} \label{appConfFac}
d\tilde{s}^2 ( \mathbb{CP}^2 )  \equiv \sin^2 \alpha \,  \Delta_1^{-1}  \,   d s^2 ( \mathbb{CP}^2 ) \; , \qquad 
\tilde{B} \equiv d\tilde{\psi} + \tilde{\sigma} \equiv X^{-1/2} \Delta_2^{-1/2}  \sin \alpha \, (d\psi + \sigma)  \; , 
\end{eqnarray}
corresponds to a deformation of the usual metric $ds_5^2  =  ds^2 ( \mathbb{CP}^2 ) + (d\psi + \sigma)^2$ on the unit radius round $S^5$ adapted to the Hopf fibration over $\mathbb{CP}^2$. 

Next, evaluating (\ref{FPeq}) on (\ref{KKSU3sectorinIIAbis}) we obtain
\be \label{SU3eqMassGravbis}
g^2 \, \Big(Xe^{2\phi-\varphi}(\partial^2_{\alpha}+5\cot\alpha\partial_{\alpha})
+ \tilde{\Box}_5  \Big)Y(y)=-M^2Y(y)\,,
\ee
where
\begin{eqnarray} \label{TildeLap}
\tilde{\Box}_5 = - \frac{1}{\sqrt{\tilde{g}}} \partial_{\tilde{m}} \Big( \sqrt{\tilde{g}} \,  \tilde{g}^{\tilde{m}\tilde{n}} \partial_{\tilde{n}} \Big)   \; , \qquad \tilde{m} ,  \tilde{n} = 1 , \ldots , 5 \; , 
\end{eqnarray}
is the scalar Laplacian of the five-dimensional metric (\ref{5Dtilde}) at constant $\alpha$. The Laplacian (\ref{TildeLap}) for the type of fibered geometries (\ref{5Dtilde}) has been computed in {\it e.g.} \cite{Hoxha:2000jf} to be
\begin{eqnarray} \label{TildeLap2}
\tilde{\Box}_5 =  \tilde{g}^{xy} \big( \tilde{\nabla}_x - \tilde{B}_x \, \partial_{\tilde{\psi}} \big)  \big( \tilde{ \nabla}_y - \tilde{B}_y \, \partial_{\tilde{\psi}} \big)  + \partial_{\tilde{\psi}}^2  \; , 
\end{eqnarray}
where $\tilde{g}^{xy}$, $x,y = 1, \ldots, 4$ are the inverse metric components corresponding to $d\tilde{s}^2 ( \mathbb{CP}^2 ) $ in (\ref{5Dtilde}). Taking into account the conformal factors in (\ref{appConfFac}), this becomes
\begin{eqnarray} \label{TildeLap3}
\tilde{\Box}_5 =   \frac{\Delta_1}{\sin^2 \alpha } \, g^{xy} \big( \nabla_x - B_x \, \partial_{\psi} \big)  \big( \nabla_y - B_y \, \partial_{\psi} \big)  +  \frac{X \Delta_2}{\sin^2 \alpha } \,  \partial_{\psi}^2  \; , 
\end{eqnarray}
where now $g^{xy}$ is the inverse Fubini-Study metric and $B = d\psi + \sigma$. Now, (\ref{TildeLap3}) can be written in terms of the scalar Laplacian $\Box_{S^5}$ on the round, unit $S^5$ by using the untilded version of (\ref{TildeLap2}),
\begin{eqnarray} \label{LapRoundS5}
 g^{xy} \big( \nabla_x - B_x \, \partial_{\psi} \big)  \big( \nabla_y - B_y \, \partial_{\psi} \big)   = \,  \Box_{S^5} - \partial_{\psi}^2
\end{eqnarray}
Finally, inserting (\ref{TildeLap3}) with (\ref{LapRoundS5}) into (\ref{SU3eqMassGravbis}), the eigenvalue equation (\ref{SU3eqMassGrav}) brought to the main text is obtained.

The graviton mass operator in (\ref{SU3eqMassGrav}) is associated to the geometry (\ref{KKSU3sectorinIIA}) corresponding to the massive IIA embedding \cite{Varela:2015uca} of the SU(3)--invariant sector of dyonic ISO(7) supergravity \cite{Guarino:2015qaa}. Accordingly, it depends on the vevs of the SU(3)--invariant scalars of the $D=4$ supergravity. More generally, we can give an expression for the graviton mass operator corresponding to the IIA embedding of the full $\cN=8$ supergravity, thus dependent on the vevs of the $\textrm{E}_{7(7)}/\textrm{SU}(8) $ scalars that enter the general background metric. The relevant ten-dimensional geometry is
\begin{eqnarray} \label{KKEmbeddingFull}
 d\hat{s}_{10}^2 = L^2 \,  \Delta^{-1} \, ds^2 ( \textrm{AdS}_4)   \, + g_{mn}  \, dy^m \, dy^n  \; ,
\end{eqnarray}
with AdS$_4$ of unit radius, so that $L^2 \,  ds^2 ( \textrm{AdS}_4)$ has radius $L$. Now, $L^2=-6 \, V^{-1}$, with $V$ the scalar potential of the full ISO(7) supergravity, and $\Delta^{-1}$ is the warp factor, dependent on the $S^6$ coordinates $y^m$, $m=1, \ldots , 6$. The inverse internal metric is \cite{Guarino:2015jca,Guarino:2015vca}
\begin{eqnarray} \label{eq:inversemet}
4 g^{-2}   \, \Delta^{-1} \,  g^{mn} = {\cal M}^{IJ \, KL} \,  K^m_{IJ} \, K^n_{KL} \; ,
\end{eqnarray}
where ${\cal M}^{IJ \, KL} $ is one of the SL(7)--covariant blocks of the $\textrm{E}_{7(7)}/\textrm{SU}(8) $ scalar matrix, and $K^m_{IJ} = 2 g^{-2} \, \mathring{g}^{mn}\mu_{[I}\partial_n \mu_{J]}$, $I=1, \ldots , 7$, are the SO(7) Killing vectors of the $S^6$ equipped with its round metric $\mathring{g}_{mn}$. Recall from the main text that $\mu^I$ are constrained coordinates in $\mathbb{R}^7$ that define the $S^6$ as the locus $\delta_{IJ} \mu^I \mu^J = 1$ and depend on the $S^6$ coordinates $y^m$. 

The spectrum of spin-2 fluctuations corresponding to the geometry \eqref{KKEmbeddingFull} is determined by inserting (\ref{eq:inversemet}) into (\ref{FPeq}). After some algebra, the resulting mass operator reads
\be
\label{KKEmbeddingFulleqn}
\tfrac{1}{4}   g^2  \,  {\cal M}^{IJ \, KL} \,  K^m_{IJ} \,\partial_m \Big( K^n_{KL}\partial_n \, Y(y)\Big)=-M^2Y(y) \; ,
\ee
The $S^6$ Killing vector ${K}^m_{IJ} \, \partial_m$ lifts to $\mathbb{R}^7$ as $2\mu_{[I}\partial_{\mu_{J]}}$. It can be seen that the $S^6$ spherical harmonics (see (\ref{KKEigenfunctionSO7})) obey
\bea
\mu_{[I}\partial_{\mu_{J]}}\, \mu^{\{I_1}\cdots\mu^{I_n\}}=C_{IJ}(J_1\cdots J_n| I_1\cdots I_n)\, \mu^{\{J_1}\cdots\mu^{J_n\}}\ , 
\eea
where we have defined
\bea
C_{IJ}(J_1\cdots J_n| I_1\cdots I_n)=-n\delta^{\{ I_1}_{[I}\delta_{J]\{J_1}\delta^{I_2}_{J_2}\cdots\delta^{I_n\}}_{J_n\}}\ .
\eea
Using these results, the general algebraic mass matrix corresponding to the differential operator on the l.h.s of \eqref{KKEmbeddingFulleqn} is deduced to be
\be
M_{I_1\cdots I_n}^{\qquad J_1\cdots J_n}=g^2 \sum_{K_1\cdots K_n}{\cal M}^{IJ \, KL}
C_{IJ}(I_1\cdots I_n | K_1\cdots K_n)C_{KL}(K_1\cdots K_n | J_1\cdots J_n) \; .
\ee
%

\section{Mass matrix trace for the $\cN=3$ solution} \label{sec:TraceMN=3}

The graviton spectrum for the massive type IIA AdS$_4$ solution \cite{Pang:2015vna} that uplifts from the $\cN=3$ SO(4) critical point \cite{Gallerati:2014xra} of dyonic ISO(7) supergravity was computed in \cite{Pang:2015rwd}. As we will now show, the trace of the graviton mass matrix in this case also follows the pattern discussed in section \ref{sec:GravMassMatD2} of the main text.

The KK graviton masses for this $\cN=3$ solution were found to be given in terms of four quantum numbers $\tilde{n}, j_F,j_V,j_{\mathcal R}$ by \cite{Pang:2015rwd}
\be \label{N=3spectrum1}
L^2M^2_{ \tilde{n} , j_F,j_V,j_{\mathcal R}} =\tfrac{1}{2} \left(2 \tilde{n} \left(4 j_F+2 j_V+5\right)+4 j_F j_V+j_F^2+7 j_F+j_V^2+5 j_V+j_{\mathcal{R}}^2+j_{\mathcal{R}}+4 \tilde{n}^2\right)
\ee
(with $n$ in \cite{Pang:2015rwd} denoted here as $\tilde n$). In terms of the SO(7) KK level $n$ of the main text, defined in this case as 
\be
n=2 \tilde{n}+2 j_F+j_V \; , 
\ee
the spectrum (\ref{N=3spectrum1}) can be rewritten as
\be \label{N=3spectrum2}
L^2M^2_{n , j_F,j_V,j_{\mathcal R}} = \tfrac{1}{2} \left(n (n+5) -3 j_F^2+j_{\mathcal{R}}^2 -3 j_F +j_{\mathcal{R}}\right). 
\ee

The branching rules for the $[n,0,0]$ representation of SO$(7)$ under the relevant SO(3) subgroups, denoted below as in \cite{Pang:2015vna}, are
\bea
[n,0,0] &\xrightarrow{\textrm{SO}(3)_L \times \textrm{SO}(3)_R \times \textrm{SO}(3)_V}&  \sum_{k= 0}^{[\frac{n}{2}]}\sum_{j=0}^{2k+1}[\tfrac{j}{2},\tfrac{j}{2},2 k+1-j]\quad, \; \text{if $n$ is odd} \\
&& \sum_{k= 0}^{[\frac{n+1}{2}]}\sum_{j=0}^{2k}[\tfrac{j}{2},\tfrac{j}{2},2 k-j]\quad,  \; \text{if $n$ is even} \\
&\xrightarrow{\textrm{SO}(3)_{F} \times  \textrm{SO}(3)_{\mathcal R} }&    \sum_{k= 0}^{[\frac{n}{2}]}\sum_{j=0}^{2k+1} \sum_{j_{\cal R}=|2k+1-\frac{3}{2}j|}^{2k+1-\frac{1}{2}j}[\tfrac{j}{2},j_{\cal R}]\quad, \; \text{if $n$ is odd} \\
&& \sum_{k= 0}^{[\frac{n+1}{2}]}\sum_{j=0}^{2k} \sum_{j_{\cal R}=|2k-\frac{3}{2}j|}^{2k-\frac{1}{2}j}[\tfrac{j}{2},j_{\cal R}]\quad, \; \text{if $n$ is even} 
\eea
where $\textrm{SO}(3)_F=\textrm{SO}(3)_L$ and $\textrm{SO}(3)_{\cal R}=[\textrm{SO}(3)_R \times \textrm{SO}(3)_V]_{\textrm{diag}} $. Now, using these branchings and the mass formula (\ref{N=3spectrum2}) we compute
\be
L^2 \; \text{tr} M_{(n)}^2=21 \, D_{n-1,9} 
\ee
for this solution. Curiously, the coefficient is the same than that for the non-supersymmetric G$_2$ solution in (\ref{eq:TrSols}). As noted in \cite{Guarino:2015qaa}, the $\cN=3$ SO(4) solution and the $\cN=0$ G$_2$ solution also have the same cosmological constant.


\section{The two scenarios of KKM and SO(8) triality} \label{sec:SO8triality}

The two scenarios of KKM \cite{Klebanov:2008vq} turn out to be related by an SO(8) triality rotation. To see this recall that, by triality, there are three different SO(7) subgroups of SO(8) (or, more precisely, Spin(7) subgroups of Spin(8)), denoted SO$(7)_v$ and SO$(7)_\pm$, such that the three inequivalent eight-dimensional representations of SO(8) decompose according to three possibilities, I, II or III:
\begin{equation} \label{eq:Triality}
\textrm{
\begin{tabular}{lllll}
I & & II & & III \\
\hline 
  $\textbf{8}_c \;
  \stackrel{\mathrm{SO}(7)_-}{\longrightarrow} \; 
   \textbf{7}+ \textbf{1}$ &  , \qquad  &
  $\textbf{8}_c \;
  \stackrel{\mathrm{SO}(7)_v}{\longrightarrow} \; 
   \textbf{8}$ &  , \qquad  &
  $\textbf{8}_c \;
  \stackrel{\mathrm{SO}(7)_+}{\longrightarrow} \; 
   \textbf{8}  $ 
   \\[5pt]
  $\textbf{8}_v \;
  \stackrel{\mathrm{SO}(7)_-}{\longrightarrow} \; 
   \textbf{8}$ &  , \qquad  &
  $\textbf{8}_v \;
  \stackrel{\mathrm{SO}(7)_v}{\longrightarrow} \; 
  \textbf{7} + \textbf{1} $ &  , \qquad  &
  $\textbf{8}_v \;
  \stackrel{\mathrm{SO}(7)_+}{\longrightarrow} \; 
   \textbf{8} $ 
   \\[5pt]
$\textbf{8}_s \;
  \stackrel{\mathrm{SO}(7)_-}{\longrightarrow} \; 
   \textbf{8}$ &  , \qquad  &
  $\textbf{8}_s \;
  \stackrel{\mathrm{SO}(7)_v}{\longrightarrow} \; 
   \textbf{8}  $ &  , \qquad  &
  $\textbf{8}_s \;
  \stackrel{\mathrm{SO}(7)_+}{\longrightarrow} \; 
   \textbf{7} + \textbf{1} $. 
\end{tabular}
}
\end{equation}
All three SO(7) subgroups of SO(8) share the same SU(3) subgroup, so we can drop the labels to denote the latter, $\textrm{SU}(3) \equiv \textrm{SU}(3)_v = \textrm{SU}(3)_\pm $. But this SU(3) commutes with a different U(1) subgroup of SO(8) inside SO$(7)_v$ and SO$(7)_\pm$. We accordingly denote these as U$(1)_v$, U$(1)_\pm$. The branching of the $\textbf{8}_v$, $\textbf{8}_+$, $\textbf{8}_-$ representations of SO(8) under each of the $\textrm{SU}(3) \times \textrm{U}(1)_v$, $\textrm{SU}(3) \times \textrm{U}(1)_+$ and $\textrm{SU}(3) \times \textrm{U}(1)_-$ subgroups can be computed by combining (\ref{eq:Triality}) with the unique decompositions (\ref{eq:7ofSO7toSO6toU3branching}) and (\ref{eq:8ofSO7toSO6toU3branching}). From these, the branching of the $\cN=8$ fields, in SO(8) representations, under the different $\textrm{SU}(3) \times \textrm{U}(1)$ subgroups may be worked out by taking appropriate tensor products and (anti)symmetrisations.  

Going through this exercise, we find that the branchings under $\textrm{SU}(3) \times \textrm{U}(1)_-$ and $\textrm{SU}(3) \times \textrm{U}(1)_v$ respectively reproduce KKM's scenarios I and II. Indeed, from (\ref{eq:Triality}), (\ref{eq:8ofSO7toSO6toU3branching}), we see that the gravitino, in the $\textbf{8}_s$ of SO(8), branches under $\textrm{SU}(3) \times \textrm{U}(1)_-$ and $\textrm{SU}(3) \times \textrm{U}(1)_v$ in the same way, yielding
\begin{equation} \label{eq:8sunderSU3U1}
\textbf{8}_s  \longrightarrow \textbf{3}_{\frac13} + \textbf{1}_{-1} + \overline{\textbf{3}}_{-\frac13} +  \textbf{1}_{+1} \; .
\end{equation}
The $\textbf{56}_s$ of SO(8), where the $\cN=8$ spin-$1/2$ fermions lie, is obtained by tensoring the $\textbf{8}_s$ with itself three times and antisymmetrising totally. It also branches under both $\textrm{SU}(3) \times \textrm{U}(1)_-$ and $\textrm{SU}(3) \times \textrm{U}(1)_v$ in the same way,
\begin{equation}  \label{eq:56underSU3U1} 
\textbf{56}_s  \longrightarrow 2 \times \textbf{1}_{+1} + 2 \times \textbf{1}_{-1} +  3 \times \textbf{3}_{\frac13} +  3 \times \overline{\textbf{3}}_{-\frac13} +  \textbf{3}_{-\frac53} +   \overline{\textbf{3}}_{\frac53} +  \textbf{6}_{-\frac13} +   \overline{\textbf{6}}_{\frac13} +  \textbf{8}_{+1} +   \textbf{8}_{-1}   \; .
\end{equation}
The identical decompositions (\ref{eq:8sunderSU3U1}), (\ref{eq:56underSU3U1}) under both $\textrm{SU}(3) \times \textrm{U}(1)$ subgroups explains why the fermion structure in both scenarios is the same.

Moving on to the bosons, the (electric) vectors lie in the adjoint of SO(8). This again decomposes under both $\textrm{SU}(3) \times \textrm{U}(1)_-$ and $\textrm{SU}(3) \times \textrm{U}(1)_v$ in the same manner,
\begin{equation}  \label{eq:28underSU3U1} 
\textbf{28}  \longrightarrow 2 \times \textbf{1}_{0} +  \textbf{3}_{\frac43} +  \overline{\textbf{3}}_{-\frac43} +  2\times \textbf{3}_{-\frac23} + 2 \times \overline{\textbf{3}}_{\frac23} +   \textbf{8}_{0}   \; ,
\end{equation}
leading to the same $\textrm{SU}(3) \times \textrm{U}(1)$ content of vectors in both scenarios. The only differences arise for the $\textbf{35}_v$ scalars and the $\textbf{35}_c$ pseudoscalars. Tensoring the $\textbf{8}_v$ and $\textbf{8}_c$ with themselves and symmetrising, we find the following decompositions under $\textrm{SU}(3) \times \textrm{U}(1)_-$:
\begin{eqnarray}  \label{eq:35vunderSU3U1ScI} 
&& \textbf{35}_v  \longrightarrow  \textbf{1}_{0}+ \textbf{1}_{-2} + \textbf{1}_{2} +  \textbf{3}_{-\frac23} +  \overline{\textbf{3}}_{\frac23} +  \textbf{3}_{\frac43} + \overline{\textbf{3}}_{-\frac43}  +  \textbf{6}_{\frac23} + \overline{\textbf{6}}_{-\frac23} +   \textbf{8}_{0}   \; , \\
\label{eq:35cunderSU3U1ScI} 
&& \textbf{35}_c  \longrightarrow 3 \times \textbf{1}_{0}  + 2\times  \textbf{3}_{-\frac23} + 2 \times  \overline{\textbf{3}}_{\frac23}  +  \textbf{6}_{-\frac43} + \overline{\textbf{6}}_{\frac43} +   \textbf{8}_{0}   \; .
\end{eqnarray}
These turn out to be swapped under $\textrm{SU}(3) \times \textrm{U}(1)_v$:
\begin{eqnarray}  \label{eq:35vunderSU3U1ScII} 
&& \textbf{35}_v  \longrightarrow  3 \times \textbf{1}_{0}  + 2\times  \textbf{3}_{-\frac23} + 2 \times  \overline{\textbf{3}}_{\frac23}  +  \textbf{6}_{-\frac43} + \overline{\textbf{6}}_{\frac43} +   \textbf{8}_{0}  \; , \\
\label{eq:35cunderSU3U1ScII} 
&& \textbf{35}_c  \longrightarrow   \textbf{1}_{0}+ \textbf{1}_{-2} + \textbf{1}_{2} +  \textbf{3}_{-\frac23} +  \overline{\textbf{3}}_{\frac23} +  \textbf{3}_{\frac43} + \overline{\textbf{3}}_{-\frac43}  +  \textbf{6}_{\frac23} + \overline{\textbf{6}}_{-\frac23} +   \textbf{8}_{0}   \; .
\end{eqnarray}
Equations (\ref{eq:35vunderSU3U1ScI}), (\ref{eq:35cunderSU3U1ScI}) reproduce the scalar and pseudoscalar charges under $\textrm{SU}(3) \times \textrm{U}(1)$ in scenario I, and (\ref{eq:35vunderSU3U1ScII}), (\ref{eq:35cunderSU3U1ScII}) do likewise in scenario II: see tables 4 and 5 of \cite{Klebanov:2008vq} (and, for the latter case, also table \ref{tab:n=0SU3U1D2} of the main text). 

The triplet of chiral superfields $\cZ^a$, $a=1,2,3$, in both infrared $\cN=2$ field theories may be thought of as descending from the vector representation, as their lowest components correspond to the coordinates $Z^a$ transverse to the M2 and D2 branes. Unlike the spinor, (\ref{eq:8sunderSU3U1}), the vector of SO(8) branches differently in scenarios I and II :
\begin{equation} \label{eq:8vunderSU3U1}
\textbf{8}_v  \stackrel{\textrm{SU}(3) \times \textrm{U}(1)_-}{\longrightarrow}  \textbf{3}_{\frac13} + \textbf{1}_{-1}  + \overline{\textbf{3}}_{-\frac13} +  \textbf{1}_{+1} \; , \qquad 
\textbf{8}_v  \stackrel{\textrm{SU}(3) \times \textrm{U}(1)_v}{\longrightarrow}  \textbf{3}_{-\frac23} + \textbf{1}_{0}  + \overline{\textbf{3}}_{\frac23} +  \textbf{1}_{0} \; .
\end{equation}
Satisfactorily enough, these branchings respectively reproduce the different charge assignments for the chirals in both $\cN=2$ field theories. In the first branching, the triplet acquires R-charge $\frac13$, in agreement with the M2-brane assignment (\ref{DeltaandRofZM2}). Similarly, in the second case the triplet has R-charge $-\frac23$, indeed reproducing the D2-brane assignment (\ref{DeltaandRofZ}). In both cases, the SU(3) singlets correspond to the transverse real directions $X^7$, $X^8$ or, equivalently, to the complex field $\cZ^4$ that is integrated out at low energy. In the M2-brane case, $\cZ^4 \sim \epsilon_{abc} \cZ^a \cZ^b \cZ^c $ \cite{Klebanov:2008vq,Klebanov:2009kp} in the infrared, again compatible with the $\pm 1$ R-charge that the first branching in (\ref{eq:8vunderSU3U1}) predicts for the singlets. In the D2-brane case, the integrated-out field $X^7$ is indeed neutral, in agreement with the second branching. In this case, though, only one of the two singlets, corresponding to $X^7$, is relevant, as $X^8$ and its associated singlet do not have a clear interpretation.

A minor point still needs to be made. We computed these branchings starting from representations of SO(8). This is obviously appropriate for the CPW solution \cite{Corrado:2001nv} as it uplifts from the SO(8) gauging, but requires some justification for the $\cN=2$ solution of massive IIA \cite{Guarino:2015jca} because SO(8) is larger than the symmetry of the problem in the latter case. In both cases, the $n=0$ KK fields fill out an $\cN=8$ superPoincar\'e supermultiplet, and this can always  be decomposed under $\textrm{SL}(8) \subset \textrm{E}_{7(7)}$ because both gauge groups, SO(8) and ISO(7), are contained in $\textrm{SL}(8)$. In the massive IIA case, the $\cN=8$ fields can be regarded as lying in representations of the semisimple factor SO$(7)$ of ISO(7). And these branch from the SL(8) representations through the intermediate SO(8) via the chain $\textrm{SL}(8) \supset \textrm{SO}(8) \supset \textrm{SO}(7)$. At this point, one still needs to decide which of the three SO(7) subgroups of SO(8) is the relevant one here. The only choice compatible with the requirement that ISO(7) (in particular, the seven translations acted upon semidirectly by its SO(7) factor) be contained in SL(8), is SO$(7)_v$. This choice fixes the $\cN=8$, $\textrm{SO}(7)$--covariant field content reported in table \ref{tab:n=0SU3U1D2}.

Finally, we note that triality also allows in principle for a scenario III, that is, an $\textrm{SU}(3) \times \textrm{U}(1)_+$--invariant vacuum. It follows from \cite{Warner:1983vz,Guarino:2015qaa} that such solution does not exist in either the SO(8) or the dyonic ISO(7) $\cN=8$ gauged supergravities. However, it could in principle exist for other $\cN=8$ gaugings, realised as an AdS, a Minkowski, or even a de-Sitter vacuum. In scenario III, the gravitini, vectors, spin-$1/2$ fermions, scalars and pseudoscalars would respectively split as
\begin{eqnarray} \label{scenarioIII} \label{ScenarioIII}
\textbf{8}_s & \longrightarrow&  \textbf{3}_{-\frac23} + \textbf{1}_{0}  + \overline{\textbf{3}}_{\frac23} +  \textbf{1}_{0} \; , \nonumber \\[5pt]
\textbf{28} & \longrightarrow& 2 \times \textbf{1}_{0} +  \textbf{3}_{\frac43} +  \overline{\textbf{3}}_{-\frac43} +  2\times \textbf{3}_{-\frac23} + 2 \times \overline{\textbf{3}}_{\frac23} +   \textbf{8}_{0}   \; , \nonumber  \\[5pt]
\textbf{56}_s & \longrightarrow& 2 \times \textbf{1}_{0} +  \textbf{1}_{-2} +  \textbf{1}_{+2} + 2\times  \textbf{3}_{-\frac23} + 2\times    \overline{\textbf{3}}_{\frac23} + 2\times  \textbf{3}_{\frac43} + 2\times    \overline{\textbf{3}}_{-\frac43}  \nonumber \\
&& + \textbf{6}_{\frac23} + \overline{\textbf{6}}_{-\frac23} +  2\times  \textbf{8}_{0}   \; ,   \\[5pt]
\textbf{35}_v & \longrightarrow & \textbf{1}_{0}+ \textbf{1}_{-2} + \textbf{1}_{2} +  \textbf{3}_{-\frac23} +  \overline{\textbf{3}}_{\frac23} +  \textbf{3}_{\frac43} + \overline{\textbf{3}}_{-\frac43}  +  \textbf{6}_{\frac23} + \overline{\textbf{6}}_{-\frac23} +   \textbf{8}_{0}    \; , \nonumber  \\[5pt]
\textbf{35}_c  & \longrightarrow  & \textbf{1}_{0}+ \textbf{1}_{-2} + \textbf{1}_{2} +  \textbf{3}_{-\frac23} +  \overline{\textbf{3}}_{\frac23} +  \textbf{3}_{\frac43} + \overline{\textbf{3}}_{-\frac43}  +  \textbf{6}_{\frac23} + \overline{\textbf{6}}_{-\frac23} +   \textbf{8}_{0}   \; ,\nonumber
\end{eqnarray}
with the same caveat as above about the SO(8) representations in the l.h.s., because the relevant gauging, if it exists, cannot be contained in SO(8). The first branching in (\ref{ScenarioIII}) indicates that the SU(3)--singlet gravitini are neutral under the would-be R-symmetry U$(1)_+$. For this reason, scenario III is not compatible with $\cN=2$ supersymmetry.


\bibliography{references}

\end{document}